\documentclass[10pt,journal,  oneside, onecolumn]{IEEEtran}

\usepackage{multirow}
\usepackage{latexsym}
\usepackage{graphicx}
\usepackage{float}
\usepackage{amsmath}
\usepackage{amsthm}
\usepackage{lipsum}
\usepackage{subfig}
\usepackage{graphicx}
\usepackage{authblk}
\usepackage{bm}
\usepackage{amsthm}
\usepackage[section]{placeins}
\usepackage{amssymb}

\usepackage{colortbl}
\usepackage{color}

\makeatletter  
\newif\if@restonecol  
\makeatother

\usepackage[linesnumbered,ruled,vlined]{algorithm2e}
\usepackage{algpseudocode}  
\usepackage{amsmath}

\usepackage{amssymb}

\usepackage{mathrsfs}
\usepackage{subfig}
\usepackage{caption}
\renewcommand{\baselinestretch}{1.0}

\begin{document}

\title{A Digital Twin Empowered Lightweight Model Sharing Scheme for Multi-Robot Systems}


\author{Kai Xiong, Zhihong Wang, Supeng Leng,~\IEEEmembership{Member,~IEEE},  Jianhua He 

\thanks{ K. Xiong, Z. Wang, and S. Leng are with School of Information and Communication Engineering, University of Electronic Science and Technology of China, Chengdu, 611731, China; and, Shenzhen Institute for Advanced Study, University of Electronic Science and Technology of China, Shenzhen, 518110, China.}
 
\thanks{J. He  is with the School of Computer Science and Electronic Engineering, Colchester, CO4 3SQ, UK.}

\thanks{The financial support of National Natural Science Foundation of China (NSFC), Grant No.62201122, and the European Union's Horizon 2020 research and innovation programme under the Marie Skłodowska-Curie grant agreement No.824019 and No.101022280.}


\thanks{The corresponding author is Supeng Leng, email: \{spleng, xiongkai\}@uestc.edu.cn}  }

\maketitle

\begin{abstract}
Multi-robot system for manufacturing is an Industry Internet of Things (IIoT) paradigm with significant operational cost savings and productivity improvement, where Unmanned Aerial Vehicles (UAVs) are employed to control and implement collaborative productions without human intervention. This mission-critical system relies on 3-Dimension (3-D) scene recognition to improve operation accuracy in the production line and autonomous piloting. However, implementing 3-D point cloud learning, such as Pointnet, is challenging due to limited sensing and computing resources equipped with UAVs. Therefore, we propose a Digital Twin (DT) empowered Knowledge Distillation (KD) method to generate several lightweight learning models and select the optimal model to deploy on UAVs. 
With a digital replica of the UAVs preserved at the edge server, the DT system controls the model sharing network topology and learning model structure to improve recognition accuracy further. Moreover, we employ network calculus to formulate and solve the model sharing configuration problem toward minimal resource consumption, as well as convergence. 
Simulation experiments are conducted over a popular point cloud dataset to evaluate the proposed scheme. Experiment results show that the proposed model sharing scheme outperforms the individual model in terms of computing resource consumption and recognition accuracy.





\end{abstract}

\begin{IEEEkeywords}
Digital Twin, Distributed Model Sharing, Knowledge Distillation, Network Calculus, Multi-Robot System.

\end{IEEEkeywords}

\IEEEpeerreviewmaketitle

\section{Introduction}

\IEEEPARstart {T}{he} advances in wireless communication, and machine learning technologies have boosted the research and development of the Industrial Internet of Things (IIoT). A multi-robot system is a typical IIoT paradigm, in which Unmanned Aerial Vehicles (UAVs) are employed to implement auto-production collaboratively without human intervention. 
It can significantly save operation costs and improve productivity \cite{{9380511Tian}}. 
Therefore, multi-robot systems have broad applications in vertical industrial fields, such as manufacturing, warehousing, and mining. 
However, like many other mission-critical IIoT applications, even minor control and operation errors in the multi-robot production lines or open-pit mining could cause substantial economic losses and safety problems \cite{{8846023Ding}, {9839387Zhou}}.
A comprehensive perception of operating environments is essential for safe and efficient multi-robot systems.

Research on environment perception for autonomous UAVs has shifted from using visual cameras to Lidar sensors \cite{8331162Hongbo}. 
The drawbacks of optical cameras are lack of depth information and sensitivity to illumination, while Lidar sensors can provide the accurate depth information and do not depend on lighting.
Many deep learning methods are proposed to percept objects from the point cloud data produced by Lidar \cite{9716056Xiaofeng}.
However, existing point cloud learning, such as Pointnet \cite{8099499Charles}, ignore the benefits of collaborative learning from multiple agents.
Diverse applications yield distinct point cloud learnings for UAV operation.
These distinct learning models can produce diversity gain by sharing knowledge with each other.
For instance, training a point cloud recognition model to identify cargo shapes in warehousing can help the scene recognition for other UAVs in autonomous piloting.
Therefore, the deep learning model sharing among UAVs can alleviate the extensive computing overheads and promote training convergence for an individual UAV.

Learning model sharing between UAVs relies on wireless communication. Unfortunately, these cloud point learning with in-depth model sizes are inappropriate for onboard training and wireless transmission. 
To alleviate the computing and communication overheads of onboard processing and model sharing, we exploit knowledge distillation (KD) to compress the trained deep model to an appropriate lightweight model for onboard processing and model sharing.
The KD method can significantly shrink the model size with little performance loss by compression.
KD has been shown to be very effective in improving the performance of a lightweight model by transferring the ‘‘dark knowledge’’ of a trained learning model, which contains the information of non-target labels \cite{2015HintonKD}. 
By contrast, the other compression methods, like network pruning quantization, and binarization, do not have this non-label learning ability \cite{9744609Sihwan}.
However, a side effect of the KD-based method is that the compression process is computationally expensive. 
Not friendly to embedded chips.


\begin{figure*}[h]
\centering
     \includegraphics[width=.95\textwidth]{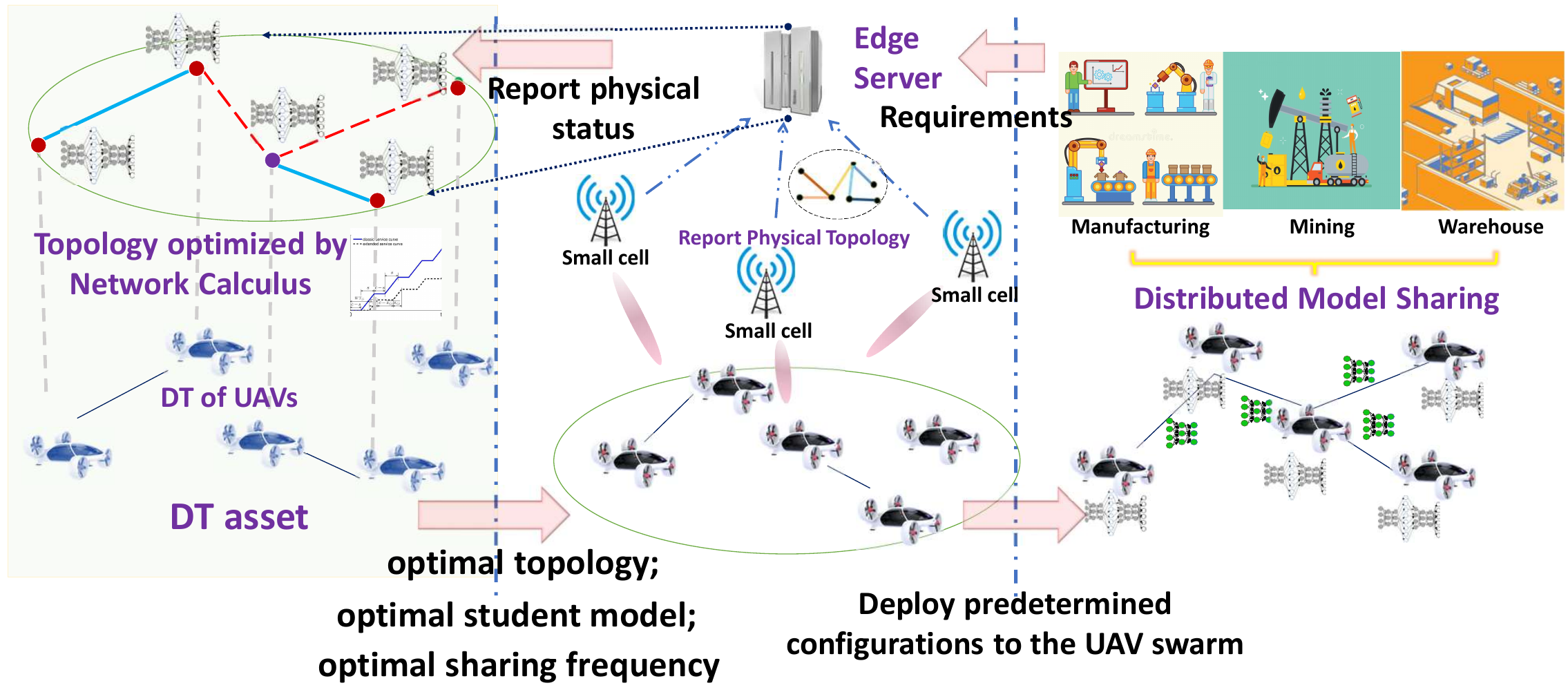} 
     \caption{DT empowered Model Sharing Framework in Multi-robot Systems. } 
\label{DT_KD_model}
\end{figure*}
Consequently, to avoid computing overheads on embedded chips and performance deterioration by lightweight model sharing, we proposed a UAV model sharing scheme with the aid of the Digital Twin (DT) technology. 
DT assets deployed in edge servers acquire the status of each UAV and perform a clone network. 
On the one hand, UAVs can offload the KD process to the DT side for computing efficiency. 
On the other hand, a DT system can inform the optimal topology configuration to the UAV side for model sharing convergence.
This is because the model sharing with different UAVs may not guarantee that all UAVs finally reach the same learnable parameters. 
It ascribes to losing the convergence of model sharing. 
In this situation, UAVs cannot fully obtain the collaboration benefits \cite{8726097XieK}. 
Therefore, an appropriate topology and resource assignment extrapolated by the DT assets will confirm the model sharing convergence.





In the DT-empowered model sharing scheme, the KD-based model training is offloaded to the DT domain.
DT asset takes advantage of the high computing capacity of edge servers to run different KD operations simultaneously.
After the KD procedure, the DT system selects an appropriate lightweight model and topological features for the UAV swarm.
This selection takes account of resource cost and convergence of model sharing.
When UAVs receive the optimal lightweight model from DT, they will retrain it for local adaptation.
In contrast, UAVs have to report their running and connection status to the DT assets for clone network refinement.
Overall, the proposed DT-based model sharing scheme improves the scene recognition efficiency of the multi-robot system with proper communication and computing overheads.

The main contributions are summarized as threefold:
\begin{itemize}

\item We propose a DT-based model sharing scheme, where the DT system comprises device DT and edge DT.
The device DT is an abstract representation of the physical UAV. The edge DT undertakes a clone network to optimize the learning model structures and the sharing network topology of the UAV swarm. 
Therefore, UAVs decouple the model distillation and network optimization from onboard chips by leveraging the computing resource and global information of edge servers. 
It reduces operation hazards and computing consumption of UAV swarms in the manufacturing context.
Additionally, the DT-based scheme further improves model sharing performance by eliminating the effect of transmission errors from physical channels.
Simulations demonstrate the advancement of the DT technology on a KD model sharing scheme. 
It produces reliable scene recognition, entitling UAVs to operate more accurately and safely.

\item We design three lightweight models of Pointnet to cater to the diverse requirements of UAVs in multi-robot systems.
Different from the traditional KD method that only generates one lightweight model and consumes computing resources on local devices, the proposed DT-based scheme enables tunable configurations in terms of the learnable parameters, model structures, and topology features.
By exploiting the computing resource of edge servers, we propose an algorithm to determine the optimal lightweight model and network topology to improve the recognition performance. 
Simulations indicate that the optimal configuration does not adopt the deepest learning model for model sharing. 
This scheme can provide a reference to address appropriate learning models on resource-limited UAVs.

\item We leverage the network calculus theory to cope with the end-to-end delay requirement for consensus convergence.
It reveals the interaction mechanism of the maximum node degree and the bandwidth of UAV-to-UAV communication as attaining a distributed consensus.
Based on the network calculus analysis and recognition requirements, we propose a consensus scheme that assigns communication and computing resources.
Simulation results exhibit that UAV network resources and topological features determine the model sharing convergence and performance.

\end{itemize}


The remainder of this paper is organized as follows. Section II designs the DT empowered distributed model sharing scheme. Section III provides the optimization of distributed model sharing. Section IV presents the simulation results and the performance discussion. Finally, we draw the conclusion in Section V.

\begin{figure}
\centering
     \includegraphics[width=.8\textwidth]{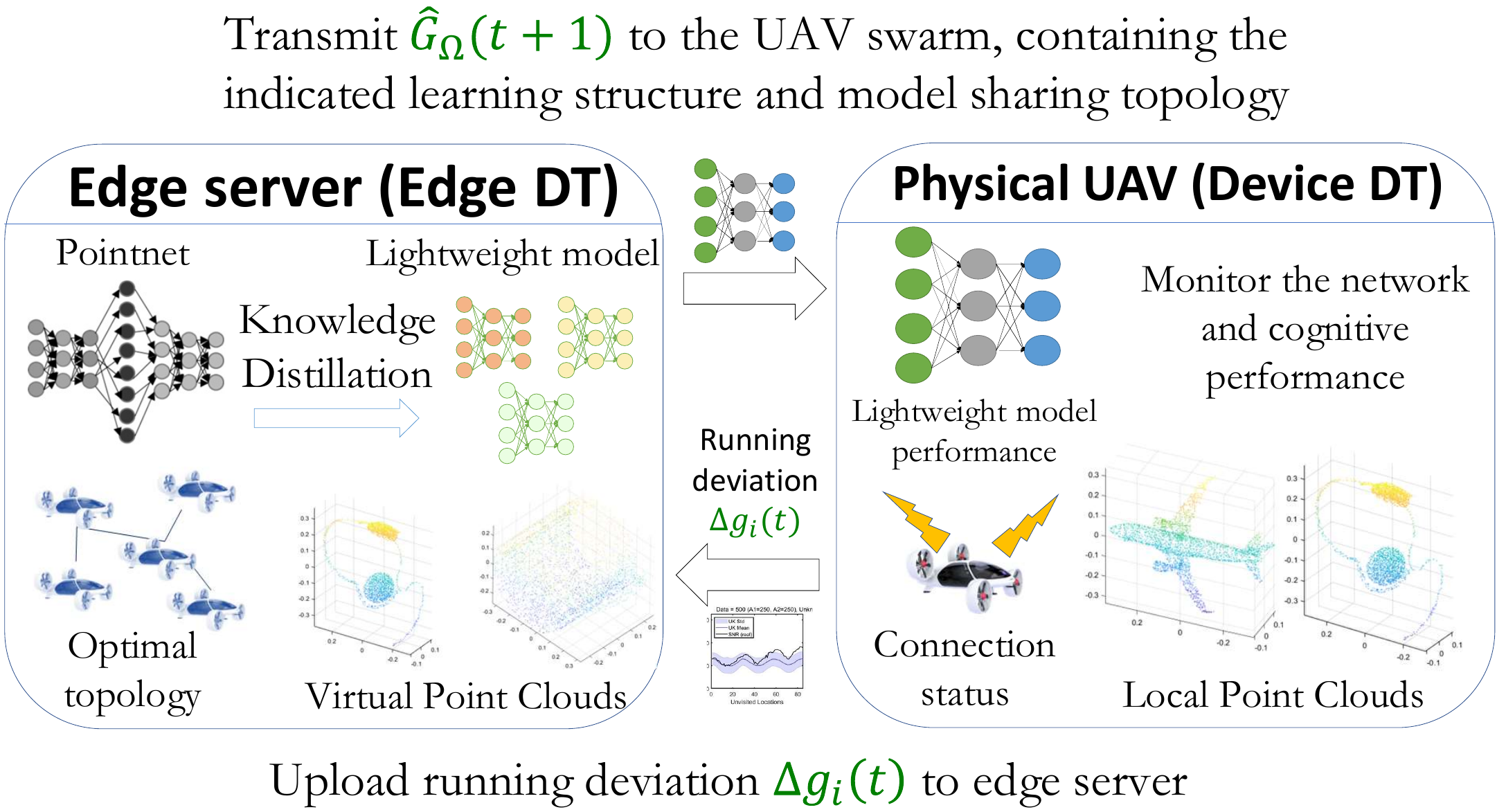} 
     \caption{DT construction and synchronization. } 
\label{DT_demo3}
\end{figure}

\section{System Model}
In this section, we present the framework of the DT-empowered model sharing and the KD-based model compression  for UAV swarms.

\subsection{DT empowered Model Sharing} 
UAVs with limited embedded sensors and chips may only partially exploit the potential of the deep learning models that requires extensive data and computing resources. 
Therefore, a DT-based scheme is proposed to yield a lightweight learning model for onboard training and efficient model parameter propagation.
A UAV obtains a lightweight learning model from the DT side, and then shares the trained model with neighbors for better recognition outcomes. This DT-based model sharing avoids the hazards caused by direct changing the physical running network.
DT assets can perform the control operations and optimization of the corresponding UAV swarm with the real-time synchronization of the physical systems.
In the DT-empowered model sharing framework, the shared model performance, size, and network topology are taken into account.

Fig.~\ref{DT_KD_model} demonstrates the proposed architecture, which consists of a physical system and synchronized digital twins.
The physical system has three major components, i.e., multiple UAVs, small cells for UAV communications and DT synchronization, and an edge server.
The UAVs gather 3-D point cloud data from onboard Lidars and then employ learning models for product recognition and autonomous pilot.
Small cells are scattered over the work region, operated for monitor and control among the UAVs and the DT assets.
The edge server can run major computing jobs such as teacher model training, lightweight model distillation, and digital twin creation for UAV swarm.

With the aid of digital twins, control and operation evaluation can be performed on the DT side.
We configure the DT model according to the reference \cite{9491087Yunlong}. 
Digital twins can be divided into device DT $DT^{dev}$ and edge DT $DT^{edge}$.
The $DT^{dev}$ is tailored to mimic the physical UAV as a summarized model. 
$DT^{edge}$ is designed to collect $DT^{dev}$ of the swarm and supply predictions and optimization to the physical system by virtual imitation and confirmation. 
In our case, the device DT is supposed to outline the UAV status, and the edge DT implies specific optimization on the mapped device DTs. This proposed DT system can facilitate DT synchronization with few communication overheads.

The assignment of device DT is to monitor and summarize the changes in UAV kinetic status, network connection status, and recognition performance.
Thus the onboard device DT of UAV $i$ is represented as, 
\begin{equation}
\begin{split}
\begin{aligned}
DT^{dev}\left(t, i\right)=\left[g_i(t), f_i(t), \Delta g_i(t) \right],
\end{aligned}
\end{split}
\label{SDAKHJAGFYUREGUSDCB}
\end{equation}

\noindent in which $g_i(t) = [g^{kin}_i(t), g^{net}_i(t), g^{mod(j)}_i(t)]$ signifies the status of the UAV kinetic, network connection, and recognition performance of the learning model $j$, respectively.
Moreover, $f_i(t) = [f^{kin}_i(t), f^{net}_i(t), f^{mod}_i(t)]$, where $f^{kin}_i(t)$ indicates the available kinetic behaviors of UAV $i$ at time $t$, such as position and acceleration.
$f^{net}_i(t)$ expresses the means of network modification at time $t$.
And $f^{mod}_i(t)$ represents the viable learning models that can be applied in UAV $i$ at time $t$.
In addition, $\Delta g_i(t) = g_i(t) - \hat{g}_i(t)$ is running deviation where $\hat{g}_i(t)$ is the prediction status at time $t$ received from the corresponding edge DT.

For the edge DT, it is a UAV swarm replicas produced by collecting several device DTs, as defined following,
\begin{equation}
\begin{split}
\begin{aligned}
DT^{edge}\left(t, \Omega \right)=\left[G_{\Omega}(t), F_{\Omega}(t),\hat{G}_{\Omega}(t+1) \right],
\end{aligned}
\end{split}
\label{SDAKHJAGFYUREGUSDCB}
\end{equation}

\noindent where $DT^{edge}\left(t, \Omega \right)$ is the edge DT corresponding to the UAV device DT $DT^{dev}(t, i), i \in \Omega$.
The set of $g_i(t)$ in the UAM swarm $\Omega$ is denoted as $G_{\Omega}(t) = [g_i(t), i\in \Omega]$. Further, $F_\Omega(t) = [f_i(t), i\in \Omega]$ represents the set of $f_i(t)$ in the group $\Omega$.
Moreover, $\hat{G}_{\Omega}(t+1)$ is the status prediction of $g(t)$ in time slot $t + 1$, which is the DT feedback to the physical system $\Omega$ for instructing future activities.

Fig.~\ref{DT_demo3} elaborates the proposed DT assets for the model sharing. 
$DT^{dev}_i$ is deployed on the UAV $i$.
On the one side, the UAV $i$ uploads its $DT^{dev}_i$ status to the edge server at intervals through wireless. 
On the other side, the edge server maintains and manipulates the gathered $DT^{dev}$ in the edge virtual renderings $DT^{edge}\left(t, \Omega \right)$.
Based on the high-performance chips of edge servers, the edge DT will extrapolate the optimal lightweight model, connection topology, and sharing frequency to instruct the physical UAV bbehaviors.

The edge DT imitates the UAV movement and experiments with manifold learning models and network configurations.
It addresses the following tasks:
i) design and select an optimal lightweight model for UAVs based on KD technology;
ii) find an appropriate topology to ensure the convergence of model sharing based on the average consensus scheme and network calculus theory;
iii) fine-tune the configuration of the model sharing to minimize communication and computing resource consumption with meeting the recognition requirement.
Wherein Generative Adversarial Network can generate the virtual training data for DT imitation in edge servers \cite{9745578Cheng}.
Through the virtual imitation and estimation, edge DT will feed back the UAV status prediction $\hat{G}_{\Omega}(t+1)$ of the following running slot $t+1$ to the UAV swarm.


In contrast, the device DT deployed on the UAV has three primary functions:
i) summarizing the local status of the UAV;
ii) maintaining device DT and synchronizing with edge DTs;
iii) onboard model training and sharing with other UAVs.
Initially, a UAV uploads its $DT^{dev}$ settings to the corresponding edge server.
Subsequently, the UAV only needs to update its running deviation $\Delta g_i(t)$ to the edge server for the DT synchronization.

The device-edge DT pairs interact through bi-directional wireless communication. 
In Fig.~\ref{DT_KD_model}, small cells collect and report the device DT information to the edge server. 
Conversely, small cells transmit the optimal topology, lightweight learning model, and sharing frequency to UAVs.
Thus, specific bandwidth resources must be reserved to sustain device DT and edge DT synchronization.

When the communication cycle exceeds the maximum allowable synchronized delay $t^{DT}$, the edge DT and physical UAV (device DT) will lose synchronization. In this scenario, the running deviation $\Delta g=null$ is sent from the device DT to the edge DT. Consequently, edge DT continues to promote virtual simulation without input information from physical devices. 
Meanwhile, the physical UAV cannot acquire feedback from the edge DT in the counterpart. 
It results in the onboard device DT only deducing the future behavior by its current status without the aid of the swarm information from edge DT. 
However, this deduction is non-optimal due to the lack of global swarm information and virtual deduction.
Once the synchronization is re-established, the device DT will transfer the current primary status $g_i(t)$ as the running deviation $\Delta g_i(t)=g_i(t)$ to the edge DT. 
After the swarm imitation and virtual trials by the edge DT, a status prediction $\hat{g}_i(t+1)\in \hat{G}_\Omega(t+1)$ at $t+1$ is given according to the upload running deviation $\Delta g_i(t)$. 
The prediction of the edge DT will instruct the UAV behavior, which rebuilds the closed-loop control between the DT system and physical entities.
 
\begin{figure*}
\centering
     \includegraphics[width=.9\textwidth]{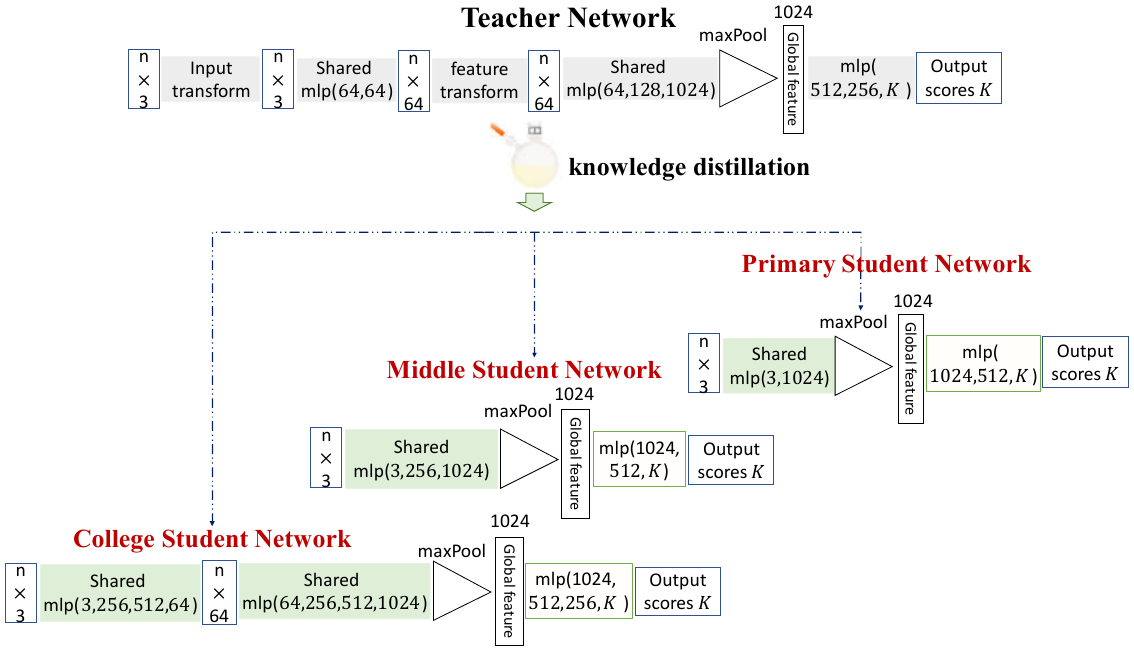} 
     \caption{{Demonstration of three KD student models for point cloud learning.}} 
\label{KD_student_model}
\end{figure*}



\subsection{Three KD Student Models}
As mentioned previously, the edge DT requires providing a suitable lightweight learning model $\hat{g}_i^{mod(j)}(t) \in \hat{G}$ to the UAVs for model sharing. 
This paper regards Pointnet learning as a teacher to produce various lightweight student models by KD compression.
Typically, the learning model to be compressed is named the teacher model, and the counterpart is called the student model.
Pointnet is classic deep learning for 3-D point cloud recognition. 
Its architecture \cite{8099499Charles} is shown at the top of Fig.~\ref{KD_student_model}.
However, the Pointnet with $14.2M$ learnable parameters is tough to train or transmit onboard in resource-limited UAVs.
Thus, this paper utilizes the KD technology deployed on edge servers to compress deep learning into a lightweight fashion.
This lightweight model can implement point cloud recognition on UAVs directly.

Unlike the existing KD method \cite{2020Empowering} that has only one student model, we propose several student models distilled from one identical Pointnet. 
Specifically, the number of learnable parameters of the three student models drops sequentially. 
It implies that student models can adapt to diverse scenarios. 
According to the order of model sizes, the three student models are named College Student Network (CSN), Middle Student Network (MSN), and Primary Student Network (PSN), respectively. 
They are illustrated in Fig.~\ref{KD_student_model}. 
Mentioned that all three student-teacher KD compressions are performed by the edge DT in that the KD process consumes lots of unaffordable computing resources with UAVs.

{
To inherit the knowledge of the teacher model, the model structure of student models should be consistent with the teacher model. 
Pointnet, i.e., the teacher model, comprises four parts: the input transform, the feature transform, the shared multi-layer perception (mlp), and the final softmax output. 
The PSN model removes the input component and feature transform. Further, it reshapes two shared mlp layers.
One shared mlp layer has $3\times 1024$ dimensions, and another degenerates into a traditional multi-layer perception with $512\times K$ dimensions. 
MSN is roughly similar to the PSN, but the size of the shared mlp layers in MSN has $3\times 256\times 512 \times 1024$, which is larger than the PSN.
CSN is different from PSN or MSN, which adds one more shared mlp block, as shown in Fig~\ref{KD_student_model}. 
The number of learnable parameters of PSN is $2.13M$, only 15\% of the teacher model. 
It bears nearly $7$ times compression. 
The parameter number of MSN is $5.27M$, about 37\% compression of the teacher model. 
And the parameter number of CSN is $6.24M$, around 44\% proportions of the teacher model.
}

The edge DT constructs the teacher-student learning models and takes advantage of the virtual data to perform the KD compression in a discrete time-slotted manner.
Suppose that the teacher and student models both generate $K$ categories scores for each of the $n$ point clouds.
The loss function of the KD is a cross-entropy between the predicted probability vector $\bm{Z}=[z_1, z_2,\dots, z_K]$ and the truth one-hot label $\bm{y}=[y_1, y_2,\dots, y_K]$ encoded by one-hot representation.
The input sample is formalized as $p\sim P_{data}(x)$, expressing the input data distribution.
Let $Z^{*SN}$ represents the logits output of student model $(Z^{PSN}/Z^{MSN}/Z^{CSN})$, and $Z^T$ indicates the logits of the teacher model.
We first get the logits $Z^T$ by training the teacher model, then distill the knowledge from the trained teacher model to supervise the student training.
Moreover, Hinton \textit{et al.} \cite{2015HintonKD} proposed a typical distillation objective to align the student logits $Z^{*SN}$ with the teacher logits $Z^T$:
\begin{equation}
\begin{split}
\begin{aligned}
min_Z \sum_{i=1}^n (1-\alpha)\mathscr{L}(\sigma(Z^{*SN}(\bm{x}_i)), \bm{y}_i) \\
+ \alpha\tau^2 \mathscr{L}(\sigma(Z^T(\bm{x}_i),\tau), \sigma(Z^{*SN}(\bm{x}_i), \tau)),
\end{aligned}
\end{split}
\label{dsfjsdhkgeriughprevnkl}
\end{equation} 

\noindent where $\tau$ signifies the distillation temperature. 
$\alpha$ is a weight parameter. 
$\bm{x}$ is the input point cloud data, and $\bm{y}$ is the truth one-hot label. 
Denote by $\mathscr{L}$ the cross-entropy to measure the gap between the teacher and student logits, defined as
\begin{equation}
\begin{split}
\begin{aligned}
\mathscr{L}(z ,y)=-\sum_{y_{j} \in \mathcal{Z}_{i}} y_{j} \log z_{i}-\sum_{y_{j} \notin \mathcal{Z}_{i}} y_{j} \log \left(1-z_{i}\right).
\end{aligned}
\end{split}
\label{cvbrjfgkocdf_v}
\end{equation}

\noindent Wherein, $\sigma$ represents the softmax layer of a deep learning network that transforms the logits $Z$ to the probability outputs, noted as,
\begin{equation}
\begin{split}
\begin{aligned}
\sigma(\bm{\bm{z}_i}) = \frac{exp(\bm{z}_i)}{\sum_j exp(\bm{z}_j)}.
\end{aligned}
\end{split}
\label{kgerrev}
\end{equation}

\noindent In addition, denote by $\sigma(z,\tau)$ the temperature augmented softmax layer, given as,
\begin{equation}
\begin{split}
\begin{aligned}
\sigma(\bm{z}_i,\tau) = \frac{exp(\bm{z}_i/\tau)}{\sum_j exp(\bm{z}_j/\tau)}.
\end{aligned}
\end{split}
\label{ktototototototv}
\end{equation}

\noindent Based on these definitions, we weightily integrate the two cross-entropy losses as Eq.~(\ref{dsfjsdhkgeriughprevnkl}), indicating the loss for student model training.

\begin{figure}
\centering
     \includegraphics[width=.65\textwidth]{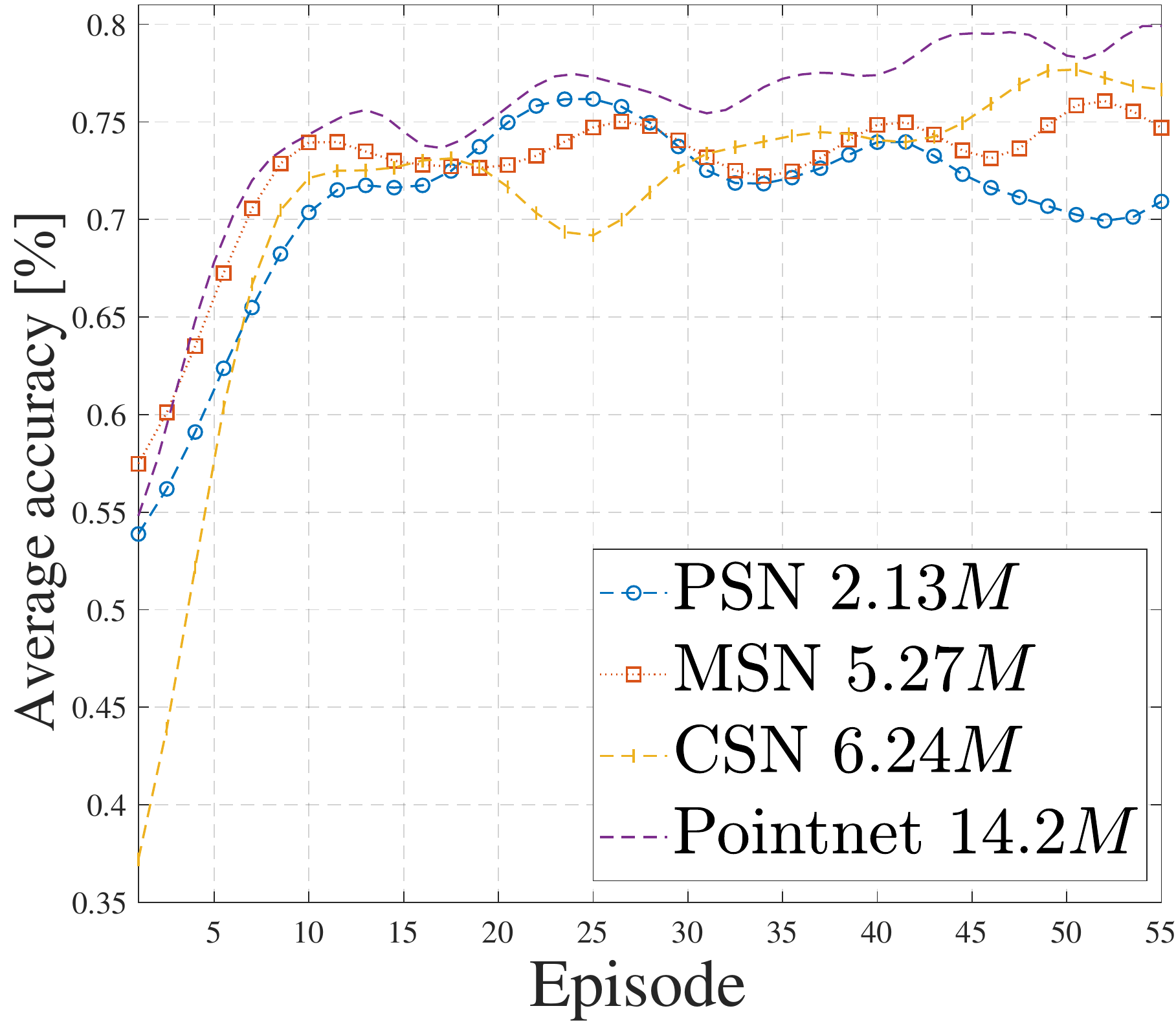} 
     \caption{Performance comparison of the teacher model with three student models. } 
\label{ind_student_teacher_performance}
\end{figure}

Fig.~\ref{ind_student_teacher_performance} depicts the recognition accuracy of the three students and teacher models in point cloud recognition.
The number of learnable parameters of each model (i.e., the model size) is marked in the legend.
The recognition accuracy of the teacher model is the upper bound of the three students since the teacher has the deepest architecture with high fitting capability.
After 15 training episodes, all student recognition accuracies are less than 10\% compared to the teacher model.

At the beginning of Fig.~\ref{ind_student_teacher_performance}, CSN has the worst accuracy in the student networks.
After 45 training episodes, it outperforms the other student models.
Because CSN has a large model size, it needs more data and training time to digest the input data and teacher knowledge. 
On the contrary, PSN and MSN can learn faster due to the simplified model structures, but the long-term training accuracy is not as good as CSN.
In addition, MSN is similar to PSN in this experiment.
The performance of the student models without model sharing can serve as the baseline for further evaluation.

\subsection{Distributed Model Sharing} 
To further improve the potential performance of multiple student models, the knowledge learned by each student model should be shared and integrated. 
This section proposes a low-complexity model sharing scheme to improve the recognition performance of student models and expand learning flexibility.
However, the parameters of the shared model between distributed UAVs may not converge to stable values. 
The phenomenon of the non-identical learnable parameters between the shared models is called disagreement. 
The disagreement manifests insufficient share in that student models can only partially draw the knowledge of all neighbors, leading to non-fully exploiting the diversity gain of model sharing.
Thus, the critical problem is to achieve a consensus agreement on model sharing between distributed UAVs. 
The consensus agreement means that the learnable parameters of the shared models generated by different UAVs are identical.
\begin{algorithm} 
	\caption{KD based Distributed Model Sharing} \label{Ag1}
	Initialize Teacher net; PSN; MSN; CSN;\\
	Train teacher model with minimizing Eq.~(\ref{cvbrjfgkocdf_v}) in DT;\\
	\For{$l:$ $1$ to $M$} 
	{
		Train PSN/MSN/CSN using Eq.~(\ref{dsfjsdhkgeriughprevnkl}) in DT; \\
		Obtain the optimal student model and network topology through Alg.~\ref{Ag2};\\
		Transmit the optimal student model and topology configuration to UAVs; \\
	\For{Traverse data batch in parallel}
	{
		Activate $i$th UAV $q$ times model sharing with neighbors $N_i$;\\
		\For{$j:$ $1$ to $q$} 
		{
			\If{UAV $i$ receives all $N_i$ neighbors updated model $h(t)$}
			{
				Update $m$ parameters (model) according to Eq.~(\ref{werpowoepaJsafdearoewIFG}); \\
			}
		}
	}
	}				
\end{algorithm}

Consider the learning model parameters $h_i$ of UAV $i$ with topology $G=(V,E)$, where each UAV can communicate with its neighbors $N_i = \{j \in V| \{i,j\} \in  E \}$ on $G=(V,E)$.
$V=\{1,2,\dots m\}$ is the set of UAVs. 
Denote by $E\in\{V,V\}$ the set of communication links.
In this scenario, Olfati-Saber \textit{et al.} \cite{1333204Olfati} has proposed the following linear dynamic method to address a distributed consensus with the UAV-to-UAV (V2V) communication delay $\epsilon$,
\begin{equation}
\begin{split}
\begin{aligned}
\dot{h}=-\operatorname{Lh}(t-\epsilon),
\end{aligned}
\end{split}
\label{ffdstrsAWEwqe}
\end{equation}

\noindent where $t$ is the time variable. 
$L=L(G)$ is the Laplacian matrix of graph $G$ that is defined as $L=D(G)-A(G)$. 
$D(G)$ is a diagonal degree matrix of $G$ with the $i$-th diagonal element $d_i=|N_i|$, and the non-diagonal elements of $D(G)$ are zero. 
$A(G)$ is an adjacency matrix with $0-1$ elements. 
Thus, the element of $L$ is
\begin{equation}
\begin{split}
\begin{aligned}
l_{i j}= \begin{cases}-1, & j \in N_i \\ \left|N_{i}\right|, & j=i\end{cases}.
\end{aligned}
\end{split}
\label{ujjjjrycsdasd}
\end{equation}

\noindent With the initial learning parameter $h_i(0) = a_i$ of model $i$, the parameters of all models asymptotically converges to the value $\bar{a}=\frac{1}{| N_i |} \sum_{i} a_{i}$.
However, this property is available while the UAVs are fully connected.
Denote by $\epsilon$ the maximum delay in all V2V communication. 
A necessary and sufficient condition for convergence of Eq.~(\ref{ffdstrsAWEwqe}) is \cite{1333204Olfati}:
\begin{equation}
\begin{split}
\begin{aligned}
\epsilon<\frac{\pi}{2 \lambda_{m}},
\end{aligned}
\end{split}
\label{werpowoepa__QKLJHDFQUAIFG}
\end{equation}
\noindent where $\lambda_{1} \leq \lambda_{2} \leq \cdots \lambda_{m}$ represent the eigenvalues of $L$. 
$\lambda_m$ is the largest eigenvalue of $L$ that can indicate the convergence rate of consensus in a network. 
According to the consensus scheme in \cite{1333204Olfati}, the model parameter integrator of UAV $i$ is designed in the following:
\begin{equation}
\begin{split}
\begin{aligned}
h_{i}(k+1, t)&=h_{i}(k, t)+\\
& \rho \sum_{j \in N_{i}} \eta_{i j}\left( h_{j}(k, t-\epsilon)-h_{i}(k, t-\epsilon) \right),
\end{aligned}
\end{split}
\label{werpowoepaJsafdearoewIFG}
\end{equation}
\noindent where $h_i$ is the learnable parameters of the UAV onboard model. 
Denote by $\eta_{i j}$ the communication link between UAV $i$ and $j$. 
As $i$ and $j$ disconnect, $\eta_{i j}$ equals to $0$. 
$\rho\in (0,1)$ is a weight depicting the influence of neighbors.

We elaborate on the workflow of the KD-based Distributed Model Sharing (KD-DMS) scheme in Alg.~\ref{Ag1}, where $M$ is the maximum episode of the KD process. 
$q$ is the sharing times in one training episode determined by the available bandwidth, DT communication overheads, and recognition accuracy of applications.
We train the teacher and student models in DT assets, then transmit the optimal student model and network topology to corresponding UAVs.
The model integrator of UAV $i$ follows Eq.~(\ref{werpowoepaJsafdearoewIFG}) in each model sharing cycle.


\section{Model Optimization by Digital Twin}
Based on the KD-DMS scheme, connected UAVs can agree upon identical parameters of the shared learning models.
However, the precondition for model sharing convergence is that the V2V communication delay is less than the consensus delay requirement.
However, the V2V delay is specified by the communication and computing resources.
Therefore, we propose an optimization scheme to guarantee consensus convergence while minimizing resource costs in the model sharing operation.
This optimization scheme is conducted on the edge DT.
Further, we leverage network calculus analysis in DT assets to determine the appropriate topological feature, sharing frequency, and communication/computing resources for efficient UAV model propagations.

\subsection{Distributed Model Sharing Optimization}
To guarantee the convergence of model sharing (i.e., the learnable parameters of all shared models are identical), the V2V communication delay $\epsilon$ should not exceed $\frac{\pi}{2 \lambda_{m}}$ based on Eq.~(\ref{werpowoepa__QKLJHDFQUAIFG}). 
Furthermore, according to Gersgorin theorem \cite{1333204Olfati}, the maximum eigenvalue of the Laplacian matrix must meet $\lambda_m\le 2 d_{max}(G)$, where $d_{max}(G)$ is the maximum node degree of topology $G$. 
Specifically, substituting $\lambda_m = 2 d_{max}(G)$ into Eq.~(\ref{werpowoepa__QKLJHDFQUAIFG}), we can get a sufficient condition for model sharing convergence,
\begin{equation}
\begin{split}
\begin{aligned}
\epsilon \leq \frac{\pi}{4 d_{\max }(G)}.
\end{aligned}
\end{split}
\label{erwtuivfyuyoa}
\end{equation}
\noindent It implies that a consensus-guaranteed topology with a specific topological feature $d_{\max }(G)$ has the corresponding delay requirement $\epsilon$.

Denote by $\delta(k)$ the size of the $k$-th student model. 
$k\in\{1,2,3\}$ expresses the PSN, MSN, and CSN, respectively.
Besides, the onboard computing cost of $k$-th student model training is $\varphi(k)$.
Let $\omega(k,q)$ denote the recognition accuracy of the $k$-th student model with sharing frequency $q$.
Moreover, the bandwidth consumption of the $k$-th student model with sharing frequency $q$ is indicated as $C(k,q)$. 
Thus the optimization model is formalized as,
\begin{equation}
\begin{split}
\begin{aligned}
\textbf{P1:} \qquad &\min\limits_{\{k, q, d_{\max }(G)\}} C(k, q)+\varphi(k) \\
\text { s.t. } \ \ \ \ \
&\text{C1:} \ D(k, q) \leq \frac{\pi}{4 d_{\max }(G)} \\
&\text{C2:} \ \varphi(k) \leq F \\
&\text{C3:} \ \omega(k, q) \geq \Theta \\
&\text{C4:} \ k \in \{1,2,3\}, \   q \ \text{and} \ d_{max}(G) \in \mathbb{N}^+
\end{aligned}
\end{split}
\label{erwqeiuidteryoa}
\end{equation}

The optimization goal is to minimize bandwidth consumption $C(k,q)$ and computing/training cost $\varphi(k)$ while meeting the consensus agreement and resource constraints. 
The first constraint \textbf{C1} is the communication delay constraint for consensus agreement of model sharing. 
In \textbf{C2}, $F$ is the available onboard computing resource of UAVs.
\textbf{C3} provides the recognition requirement $\Theta$ of the UAV application. 
The value range of $k$, $q$, and $d_{max}(G)$ are set into \textbf{C4}, where $\mathbb{N}^+$ is the set of positive integer.

The optimization variables of $P1$ involve the student model $k$, sharing frequency $q$, and topological feature, i.e., the maximum node degree $d_{\max }(G)$. Note that $d_{\max }(G)$ only affects the convergence of the model sharing agreement and has no consequence on the final recognition accuracy.


\subsection{Network Calculus Analysis}
The subsection reveals the relation between a delay upper bound $D(k, q)$ of \textbf{C1} and resource consumption of communication and computing. This estimation is performed in the edge DT.
First, we sort out the resource consumption caused by DT synchronization.
The DT synchronization comprises the running deviation $\Delta g$ upload from UAVs to edge servers; and the status prediction $\hat{G}_{\Omega}(t+1)$ download from edge servers to UAVs.
We only consider the upload and DT processing in the DT synchronization \cite{9491087Yunlong} to simplify the system analysis. 

Wherein the upload delay can be written as $\frac{\Delta g}{C^{DT}}$, in which $C^{DT}$ is the reserved bandwidth for DT synchronization.
Moreover, the DT synchronization delay is determined by the edge server processing capacity $\upsilon_{DT^{edge}}$ and the value of running deviation $\Delta g$. 
It can be written as $\frac{\chi \Delta g}{\upsilon_{DT^{edge}}}$, where $\chi$ is a metric to measure the DT computing complexity.
Assuming the maximum arrowed DT synchronization delay is $t^{DT}$, it must meet $t^{DT} \ge \frac{\Delta g}{C^{DT}} + \frac{\chi \Delta g}{\upsilon_{DT^{edge}}}$. 
Therefore, we can obtain the reserved bandwidth for DT synchronization as,

 \begin{equation}
\begin{split}
\begin{aligned}
C^{DT} = \frac{\Delta g  {\upsilon_{DT^{edge}}}    }{ t^{DT} {\upsilon_{DT^{edge}}}   -  {\chi \Delta g}   }. 
\end{aligned}
\end{split}
\label{feigurwe_iogws}
\end{equation} 

Hereafter, according to the network calculus, we acquire a function with the reserved bandwidth $C^{DT}$, the student model $k$, sharing frequency $q$, and the maximum node degree $d_{max}(G)$, in the model sharing convergence condition.

In network calculus theory, an arrival process $A(s,t) = A(t) - A(s)$ defines a cumulative number of the input network traffic of a UAV in the time interval $(0,t]$ \cite{Rizk6868978}. 
There are $|N_i|$ neighbors of UAV $i$. 
$\delta(k)$ represents the volume of the shared $k$-th student model parameters. 
And the transmitted rate $\mathcal{E}$ is used to maintain the basic connection with neighbors. 
Due to the $(\epsilon, \sigma)$-upper constraint, we have $A(t)-A(s) \leq \mathcal{E}(t-s)+q \delta(k) \frac{t-s}{t}$ \cite{Jiang2008Stochastic}. 
Thus, the arrival curve of UAV $i$ for sharing the model to neighbors is
\begin{equation}
\begin{split}
\begin{aligned}
A_{k, q}(t)=\mathcal{E} t+q \delta(k).
\end{aligned}
\end{split}
\label{_sdfasior_oa}
\end{equation} 
 
\noindent Moreover, the channel service curve $S$ for model sharing can be written as
\begin{equation}
\begin{split}
\begin{aligned}
S(s, t)=(C-C^{DT} ) \cdot(t-s),
\end{aligned}
\end{split}
\label{asioasdasdaw}
\end{equation} 
\noindent where $C$ is the total bandwidth consumption. 
$C-C^{DT} $ indicates the available bandwidth for model sharing that has removed the DT bandwidth consumption.
We assume that all neighbors $N_i$ of UAV $i$ contain the same arrival curve. 
The number of neighbors participating in the model sharing is $|N_i|$.
Upon the Leftover Service theorem \cite{Jiang2008Stochastic}, the channel service curve for an individual UAV is
\begin{equation}
\begin{split}
\begin{aligned}
S^{\prime}(t)=(C-C^{DT} ) \cdot t-\left(|N_i|-1\right)[\mathcal{E} \cdot t+q \delta(k)].
\end{aligned}
\end{split}
\label{erskdalfjhw}
\end{equation} 

\noindent Based on the upper bound delay analysis \cite{Jiang2008Stochastic}, the end-to-end maximum delay is $\min \left\{\omega \geq 0: \max _{\epsilon \in[0, t]}\{A_{k,q}(\epsilon, t)-S^{\prime}(\epsilon, t+\omega)\} \leq 0\right\}$. 
Via several algebraic operations, we get the upper bound of the V2V communication delay as, 
\begin{equation}
\begin{split}
\begin{aligned}
D(k, q)=\frac{|N_{i}| q \delta(k)}{C -\left(|N_{i}|-1\right) \mathcal{E}-C^{DT}}.
\end{aligned}
\end{split}
\label{__ripuewrfsdhuaifgw}
\end{equation} 

\noindent Therefore, the upper bound delay $D(k, q)$ can be regarded as the delay requirement $\epsilon$ for model sharing convergence as shown in Eq.~(\ref{erwtuivfyuyoa}).
And substituting $D(k, q) = \epsilon \leq \frac{\pi}{4 d_{\max }(G)}$ into Eq.~(\ref{__ripuewrfsdhuaifgw}), the requirement of bandwidth for model sharing convergence holds that
\begin{equation}
\begin{split}
\begin{aligned}
C(k, q) \geq \frac{4 d_{\max }(G)}{\pi} |N_i| q \delta(k)+\left(|N_i|-1\right) \mathcal{E} + C^{DT} .
\end{aligned}
\end{split}
\label{qweqweo0998aifgw}
\end{equation} 

\noindent Hence to minimize the bandwidth consumption, $C(k,q)$ should accept the equality in Eq.~(\ref{qweqweo0998aifgw}).
Therefore, we can revise the optimization problem \textbf{P1} as,
\begin{equation}
\begin{split}
\begin{aligned}
\textbf{P2:} \ \min\limits_{\{k, q, d_{\max }(G)\}} &\frac{4 d_{\max }(G)}{\pi} |N_i| q \delta(k)+\left(|N_i| -1\right) \mathcal{E}+\varphi(k) \\
\text { s.t. } \ \ \ 
&\textbf{C2}, \ \textbf{C3}, \ \textbf{C4}.
\end{aligned}
\end{split}
\label{erwqeiuidteryoa}
\end{equation}

\noindent Since $C^{DT}$ is irrelated to $\{k, q, d_{\max }(G)\}$, it can be considered as a constant in the optimization.
The exact value of $C^{DT}$ does not impact the final optimal loss. 
Thus, we remove $C^{DT}$ in the objective of $P2$. 
Moreover, the transmitted data volume $\delta(k)$ and computing cost $\varphi(k)$ are monotonically increasing with $k$. 
It means that minimizing $k$, $q$ and $d_{\max }(G)$ with constraints can achieve the goal of \textbf{P2}.

\begin{algorithm} 
	\caption{DT based Optimal Configuration Search} \label{Ag2}
	Initialize $k_0=1$; $k_1$ is the max number of $k$ with $\varphi(k_1)\le F$; $q_0=1$; $q_1$ is the max number of $q$; $d_{max}(G)=2$\\
	
\While{$ \omega(k_0, q_0) \le \Theta$}
{
	\While{$k_0 \neq k_1$}
	{
		Set $k_m=\lfloor \frac{k_1 - k_0}{2}\rfloor + k_0$; \\
		\If {  $\omega(k_m, q_1) > \omega(k_0, q_1)$ and $\omega(k_m, q_1) > \omega(k_0, q_1)$ }
			{
				$k_0 =\lfloor \frac{k_m - k_0}{2} \rfloor$; \
				$k_1 =\lfloor \frac{k_1 - k_m}{2} \rfloor$;\\
			}
			
		\ElseIf {  $\omega(k_1, q_1) \ge \omega(k_m, q_1) \ge \omega(k_0, q_1)$ }
			{
				$k_0 = k_m$; \
				$k_m = \lfloor \frac{k_1 - k_m}{2} \rfloor$;
				
			}
		
		\ElseIf {  $\omega(k_0, q_1) \ge \omega(k_m, q_1) \ge \omega(k_1, q_1)$ }
			{
				$k_1 = k_m$; \ 				
				$k_m = \lfloor \frac{k_m - k_0}{2} \rfloor$;
				
			}
	}

	\While{$q_0\le q_1$}
	{
		\If{$ \omega(k_0, q_0) \ge \Theta$}
		{
			\Return $(k_0,q_0, d_{max}(G) )$ 
		}
		
		$q_0 = q_0+1$;		
	}
	
	$d_{max}(G)=d_{max}(G) + 1$;\\
}

\Return $(k_0,q_0, d_{max}(G) )$ 
				
\end{algorithm}

However, the learning model's recognition accuracy $\omega(k, q)$ is a non-analytical and non-differentiable function of $k$ and $q$. 
When the search space of $k$-$q$ pairs is small, $\omega(k, q)$ can be enumerated by the edge DT and acquired through the look-up table method.
Tab.~\ref{table_1} depicts the recognition accuracy $\omega$ with different $(k,q)$.
\begin{table}[htb]
\begin{center}   
\caption{recognition accuracy with different $q$-$k$ pairs.}  
\label{table_1} 
\begin{tabular}{|c|c|c|c|}   
\hline   \textbf{$\omega(k, q)$} & \textbf{PSN $k=1$} & \textbf{MSN $k=2$} & \textbf{CSN $k=3$} \\   
\hline   Frequency $q=1$ & 0.780 & 0.821 & 0.769 \\ 
\hline   Frequency $q=2$ & 0.734 & 0.804 & 0.787 \\  
\hline   Frequency $q=3$ & 0.817 & 0.923  & 0.872 \\      
\hline   Frequency $q=4$ & 0.890 & 0.935  & 0.902 \\  
\hline   Frequency $q=5$ & 0.739 & 0.930  & 0.895 \\  
\hline   
\end{tabular}   
\end{center}   
\end{table}
The experiment configuration refers to the settings of the Simulation section.
The recognition accuracy of PSN does not increase with the sharing frequency $q$, in that PSN has a simple model structure without the learning capability to extract data features.
UAVs with PSN thereby share an inaccurate model, which sparks model contamination and deteriorates the sharing performance.
To be rough, the recognition accuracy of MSN and CSN rises with $q$. 
Their recognition accuracies outperform that of PSN.
Meanwhile, the performance of CSN is worse than that of MSN since CSN may suffer overfitting.
When the accuracy requirement is $\Theta=0.9$, only $(k=2, q=3,4,5)$ and $(k=3, q=4)$ qualify.
Moreover, the minimum bandwidth and computing cost is obtained at $(k=2, q=3)$.

Through the algebraic transform of Eq.~(\ref{qweqweo0998aifgw}), the model sharing frequency $q$ is constrained by,
\begin{equation}
\begin{split}
\begin{aligned}
q \le \frac{[C - (|N_i|-1)\mathcal{E}-C^{DT}]\pi}{4 d_{max}(G)|N_i|\delta(k)}.
\end{aligned}
\end{split}
\label{esdfqwefuioosdayoasfaa}
\end{equation}

\noindent In our simulation configuration, the maximum node degree $d_{max}(G)=3$ and the number of neighbors $|N_i| = 6$.
According to the above Eq.~(\ref{esdfqwefuioosdayoasfaa}), we can derive the optimal $q=3$.


Referring to Tab.~\ref{table_1}, we find that $\omega(k,q)$ increases first and then decreases with $k$. 
The reason is that $\omega(k,q)$ raises with $k$ till the $k$-th student model encounters overfitting.
Therefore, a binary search approach is proposed to explore the optimal $k$ to maximize $\omega(k,q)$. 
When the space of $k$-$q$ pairs becomes large, we can leverage $q$, $k$, and $d_{max}(G)$ properties to quickly solve \textbf{P2}.
Since the optimal $q$ is usually small due to the limited bandwidth, we investigate the value of $q$ from $q=1$ in Tab.~\ref{table_1}.
In addition, $d_{max}(G)$ is at least $2$ in a fully connected topology with over $3$ nodes. 
Further, our analysis reveals that a smaller $d_{max}(G)$ has less bandwidth cost, as indicated in Eq.(\ref{qweqweo0998aifgw}).
In summary, we propose a search scheme for model sharing configuration, as shown in Alg.~\ref{Ag2}.
The time complexity of the proposed approach is $O(n\log{(n)})$ in that the optimal sharing frequency $q$ is less than $5$ in most cases.
 
Alg.~\ref{Ag2} is implemented on the edge server for searching the optimal model sharing configuration.
The edge DT can manipulate the maximum node degree of the clone network to optimize $\textbf{P2}$ without practical network alteration and deployment costs.
Eventually, the edge DT provides the most suitable student model, communication topology, and sharing frequency in the next time slot to steer the physical model sharing arrangement.

\section{Performance Evaluation} 
To evaluate and compare the proposed DT-empowered KD-DMS algorithm applied to model sharing, we investigate the different student models, topologies, and sharing frequencies on a popular point cloud dataset, i.e., ShapeNet.
This dataset consists of $16681$ samples belonging to $16$ common categories \cite{{3326362YueWang},{9302662Xiaofeng}}.
$2500$ point clouds with coordinates are uniformly sampled as input for each UAV.
To facilitate learning diversity, we divide the dataset into four equal parts, and each UAV randomly takes three of the four parts as its database.
The training dataset for each UAV is independently sampled from its database, with a sampling probability of $0.5$, to imitate the distinction of UAV data collections. 
Furthermore, the test set is based on the original split test set in ShapeNet.
Our proposed KD-DMS algorithm is run on a computer with NVIDIA GeForce RTX 2060 GPU and coded with PyTorch.
The Adam optimizer with a $0.001$ learning rate is set to model training. The batch size is $32$, and the total training episode is $60$. Some important simulation parameters are listed in Tab.~\ref{table_2}.

\begin{table}[htb]
\begin{center}   
\caption{Simulation parameters}  
\label{table_2} 
\begin{tabular}{|c|c|c|c|}   
\hline   learning rate & 0.001 & sharing frequency & 5 \\ 
\hline   batch size & 32 & sampling probability & 0.5\\  
\hline   Number of category & 16 & data volume of each UAV & 2500 \\  
\hline   KD Temp $\tau$ & 20 & KD weight $\alpha$ & 0.5 \\     
\hline
\end{tabular}   
\end{center}   
\end{table}

\begin{figure}
\centering
     \includegraphics[width=.65\textwidth]{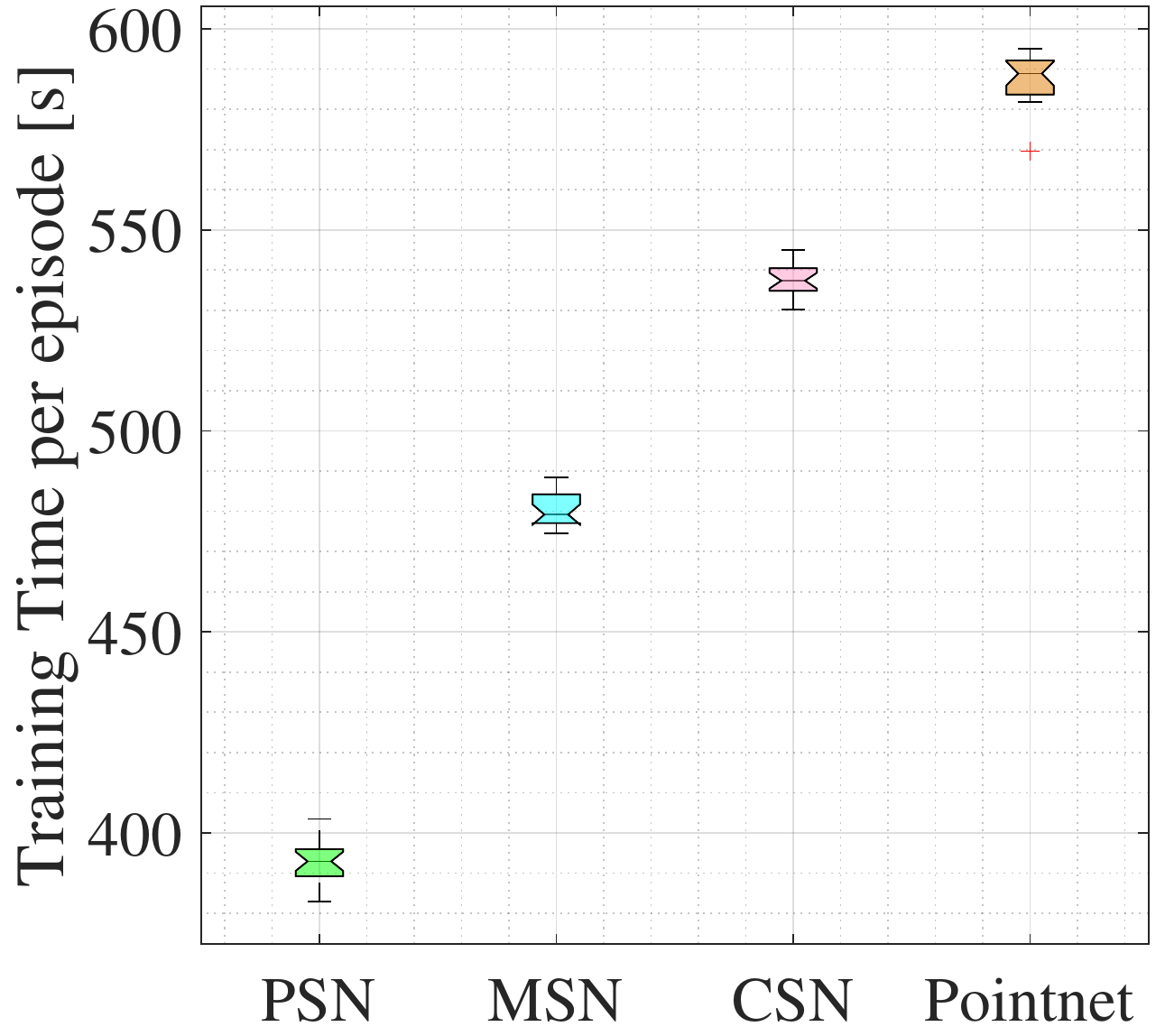} 
     \caption{Time consumption of different models per episode. } 
\label{TIMEcompare}
\end{figure}

\begin{figure}
\centering
     \includegraphics[width=.5\textwidth]{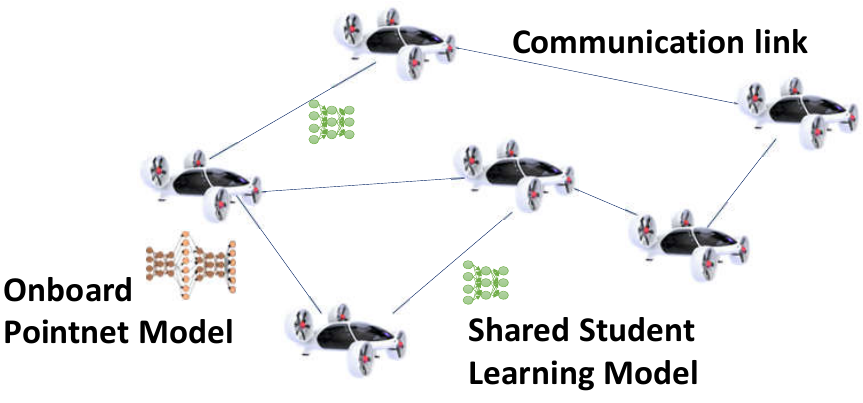} 
     \caption{{ Model Sharing Topology with the maximum node degree $d_{max}(G) = 3$, where the onboard Pointnet of flying cars generates a lightweight student model by the knowledge distillation, then the flying cars transmit the student model with neighbours according to the topology. }} 
\label{KD_DMS_Topo}
\end{figure}

 \begin{figure}
\centering
     \includegraphics[width=.65\textwidth]{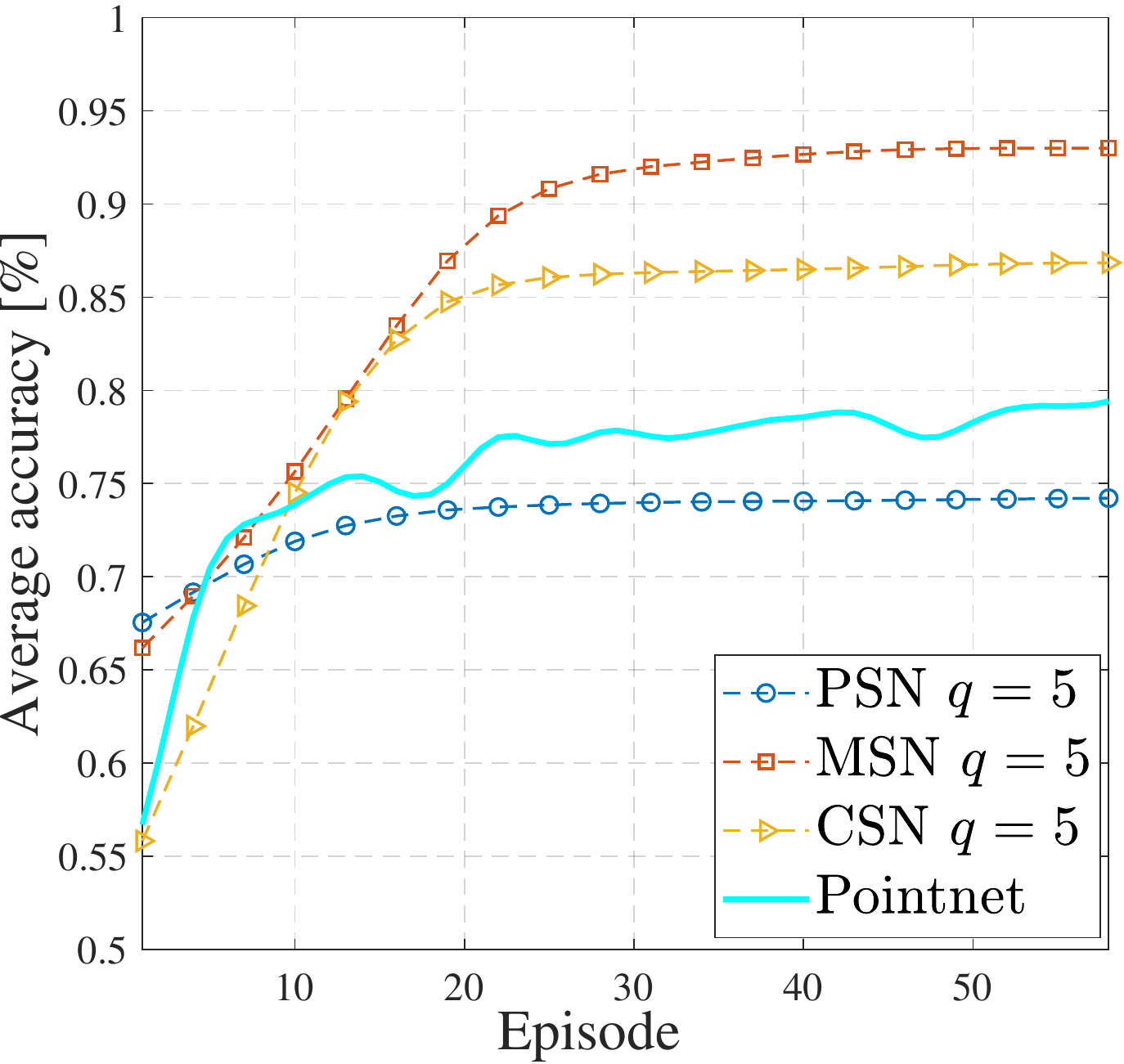} 
     \caption{Performance of KD-DMS with different learning models. } 
\label{KD_DMS_Compare}
\end{figure}

 \begin{figure}
\centering
     \includegraphics[width=.65\textwidth]{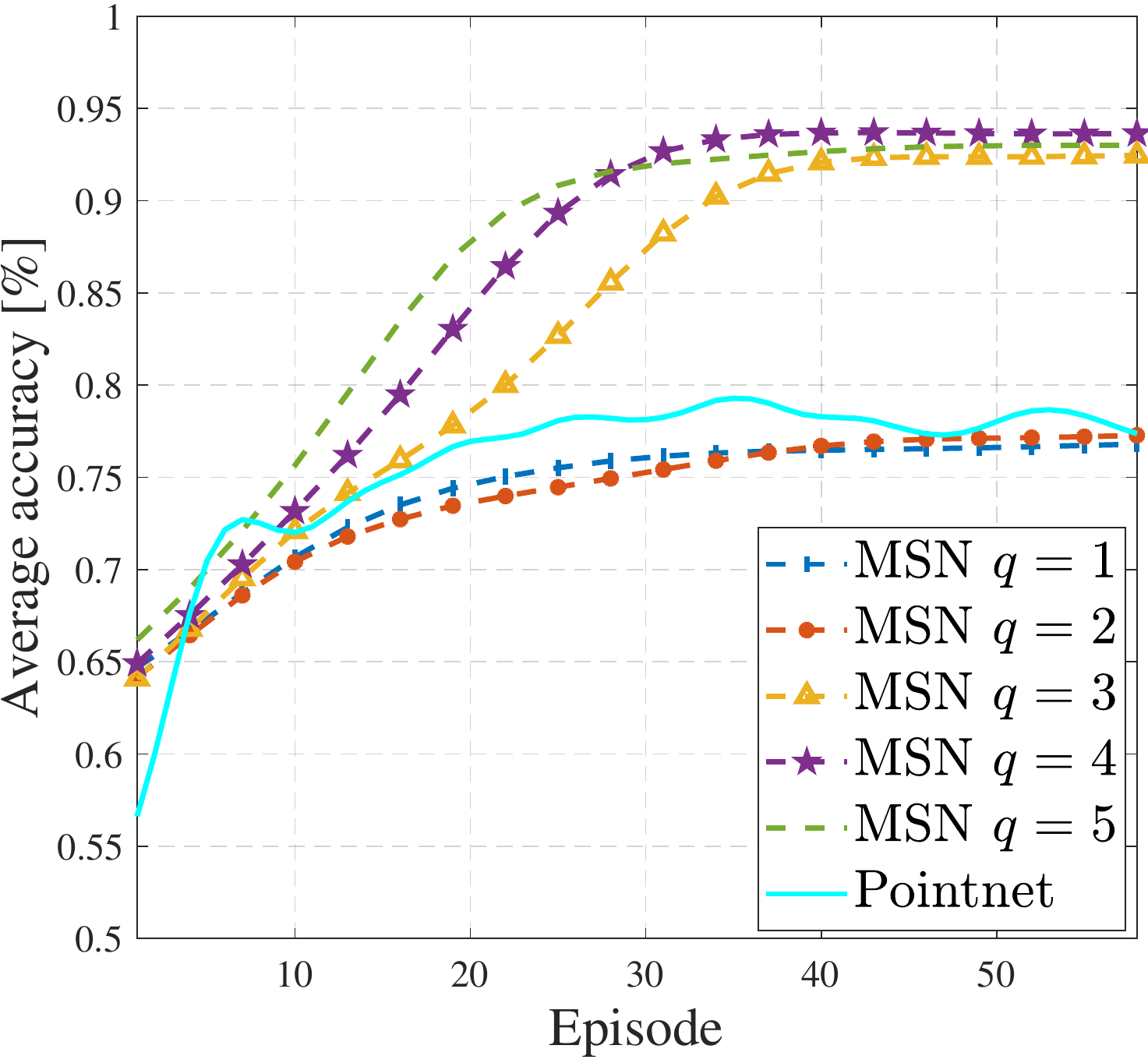} 
     \caption{Performance of MSN with different $q$. } 
\label{MSN_q_Teach_compare}
\end{figure}


Fig.~\ref{TIMEcompare} is a box plot of the time consumption versus different models, in which the $y$-axis represents the training time. 
The teacher model, Pointnet, has the maximum training time. 
The time consumption of PSN is only 66\% of Pointnet. 
Moreover, MSN and CSN are 81\% and 91\% of Poinetnet in the time consumption, respectively. 
Thus, all three KD models substituting the Pointnet can accelerate the onboard training process of UAVs.

\begin{figure*}
\centering
\subfloat[PSN.]{\label{dsfjhjklhalllokoko}{\includegraphics[width=0.31\linewidth]{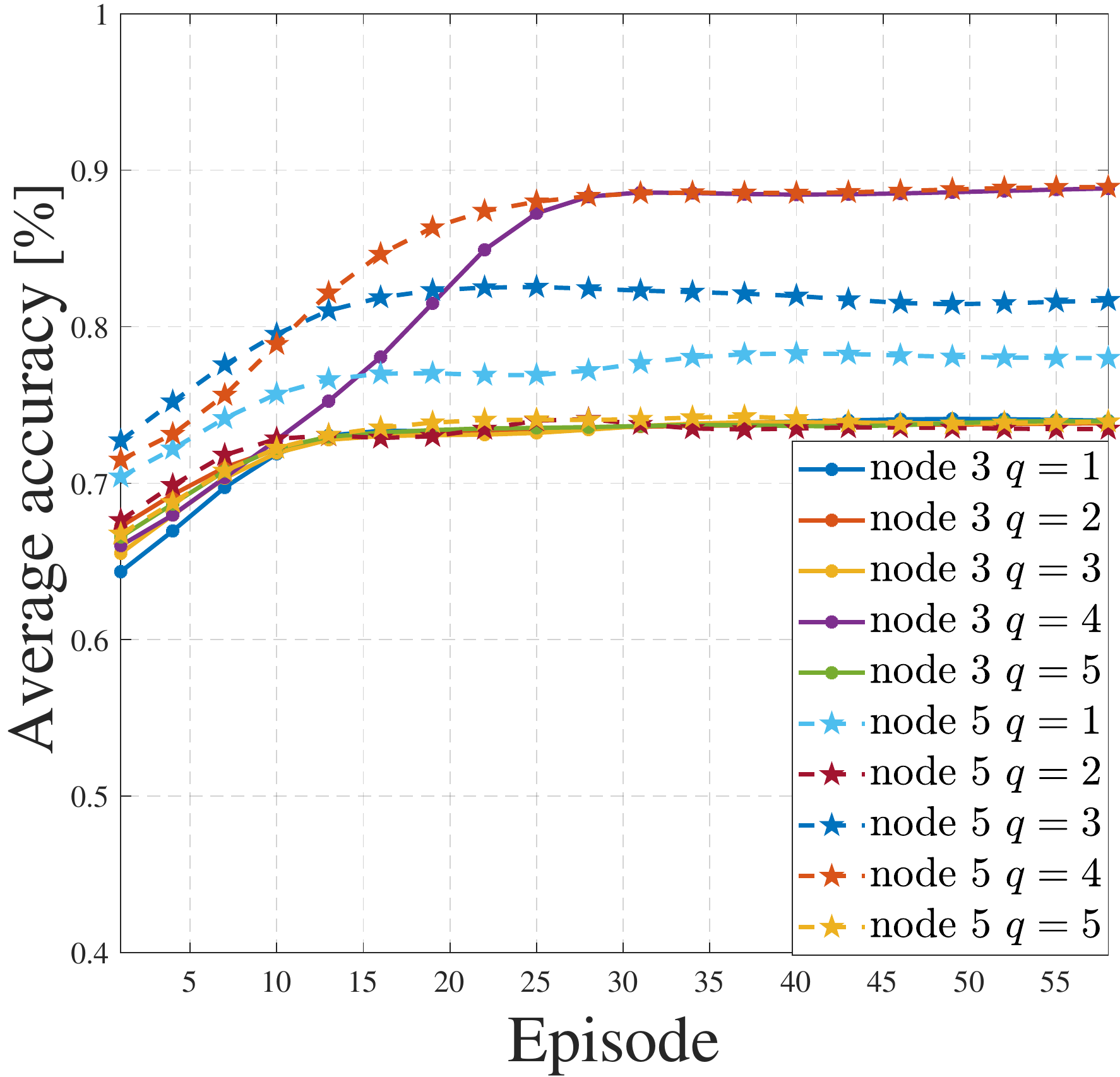}}} \hfill
\subfloat[MSN.]{\label{kjhkjhkjasappp}{\includegraphics[width=0.31\linewidth]{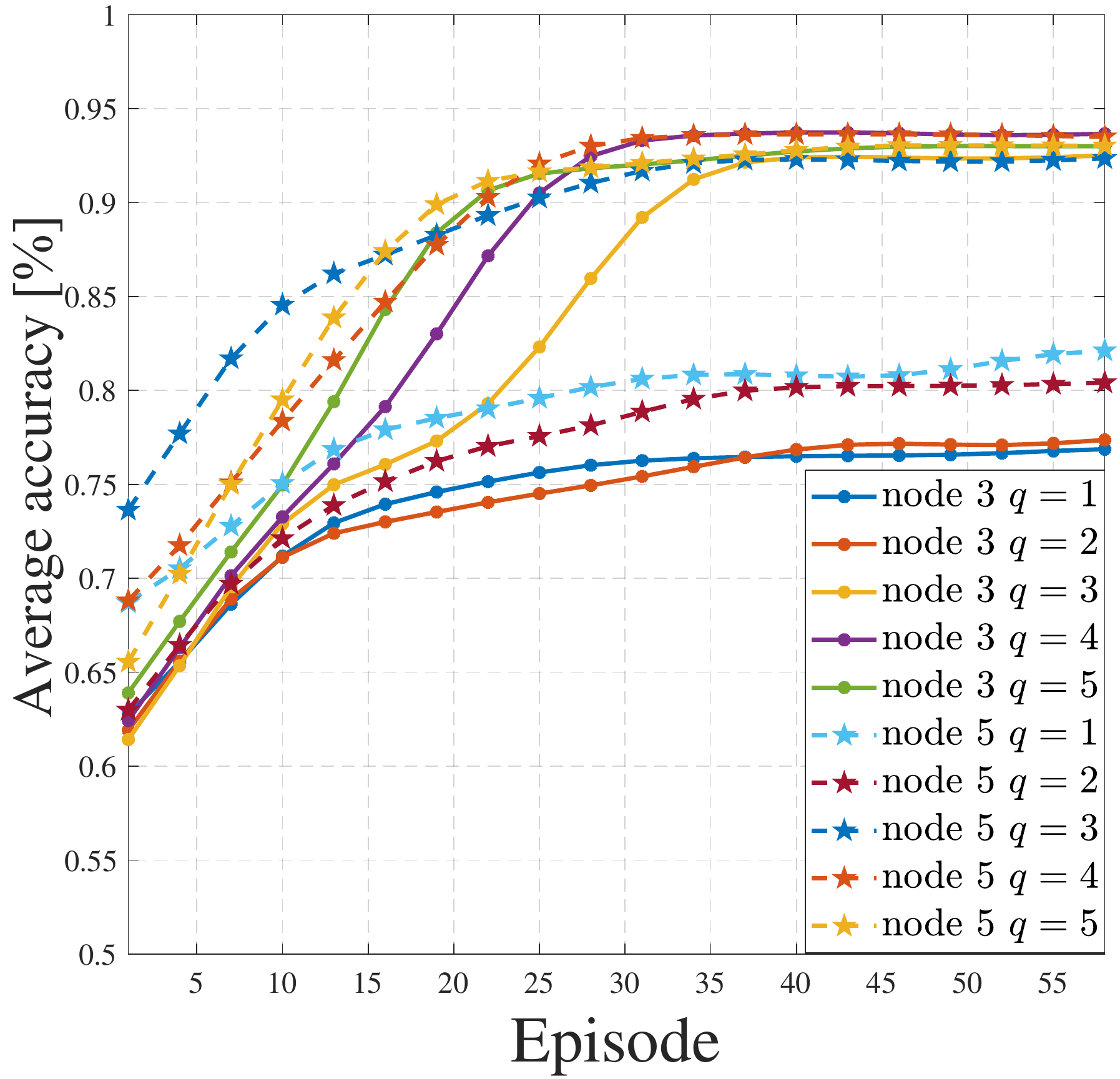}}} \hfill
\subfloat[CSN.]{\label{vvkvkvkvkvkvvsadada}{\includegraphics[width=0.338\linewidth]{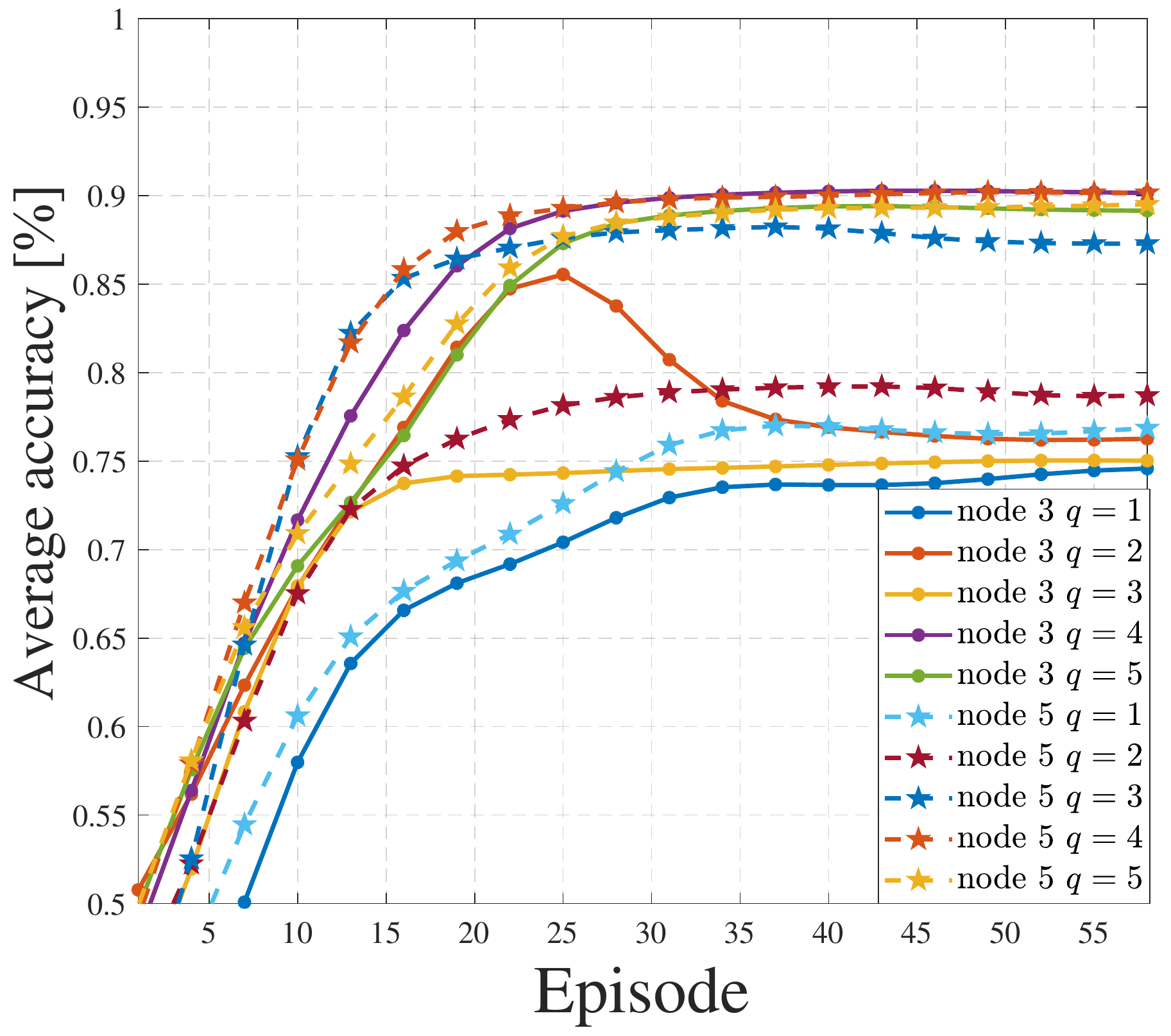}}}\hfill
\caption{Performances of different nodes (UAVs) with sharing frequency $q$ in various student models.}
\label{fig_sim_asdkjajldws}
\end{figure*}

{
Fig.~\ref{KD_DMS_Topo} illustrates the corresponding communication topology for model sharing.
We present the recognition accuracy of three KD models with model sharing compared to the Pointnet without model sharing in Fig.~\ref{KD_DMS_Compare}. 
As illustrated in Fig.~\ref{KD_DMS_Compare}, the proposed KD-DMS with $q=5$ in MSN and CSN outperforms the Pointnet without model sharing in terms of recognition accuracy after ten episodes.}
Through the model sharing in Eq.~(\ref{werpowoepaJsafdearoewIFG}), student models can integrate trained learning parameters of neighbors on different datasets.
The model sharing on distinct datasets provides diversity training gain. 
Moreover, the recognition of KD-DMS with MSN model completely surpasses that of KD-DMS with CSN model.
The possible reason is that CSN involves more learning parameters than MSN, which is more likely to incur an inconsistent/error transmission problem.


\begin{figure}
\centering
\subfloat[MSN recognition performance.]{\label{fig_sim_noCrossEntropyScore_0}{\includegraphics[width=0.65\linewidth]{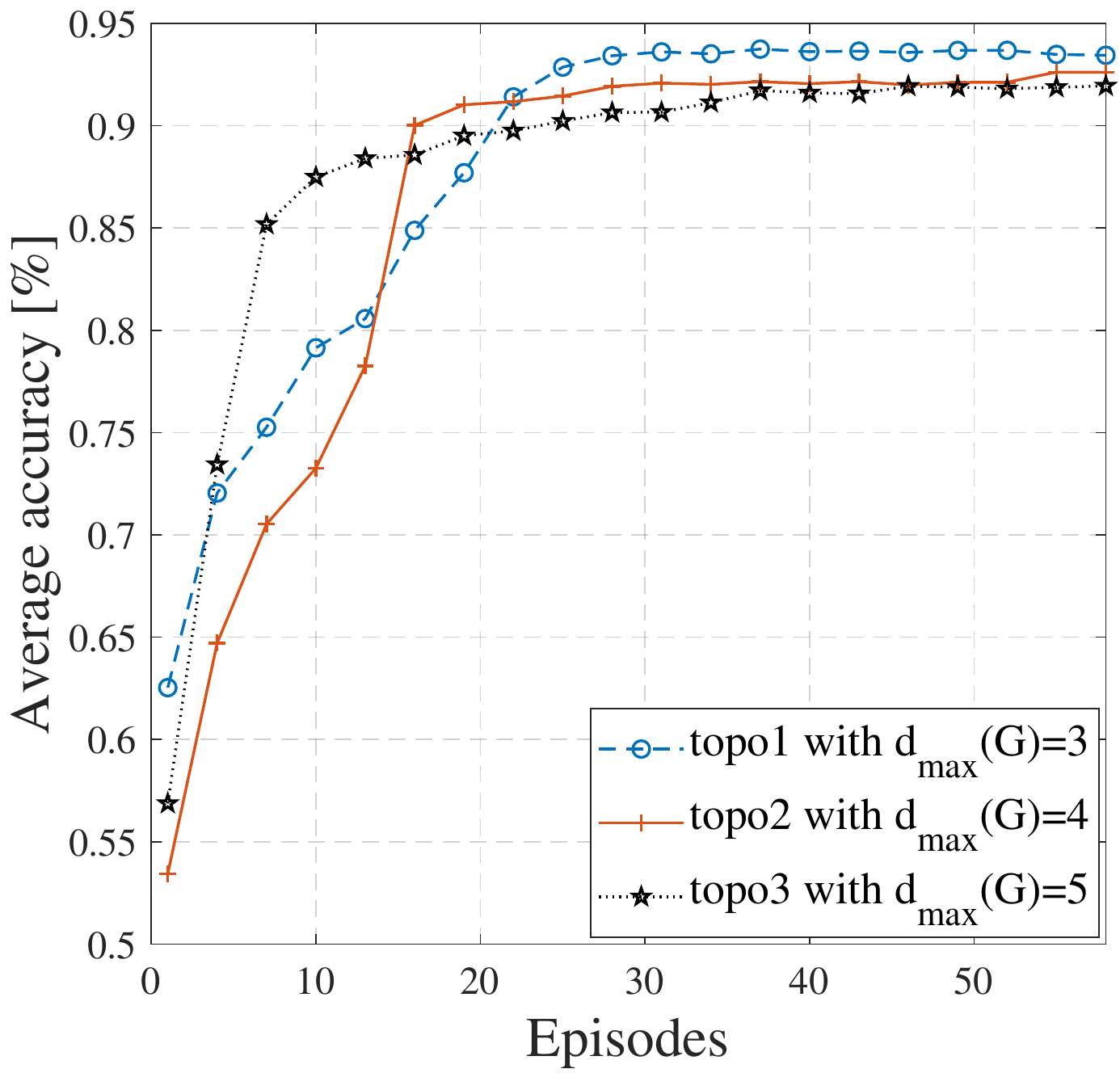}}}\hfill
\subfloat{\label{fig_sim_noCrossEntropyScore}{\includegraphics[width=0.32\linewidth]{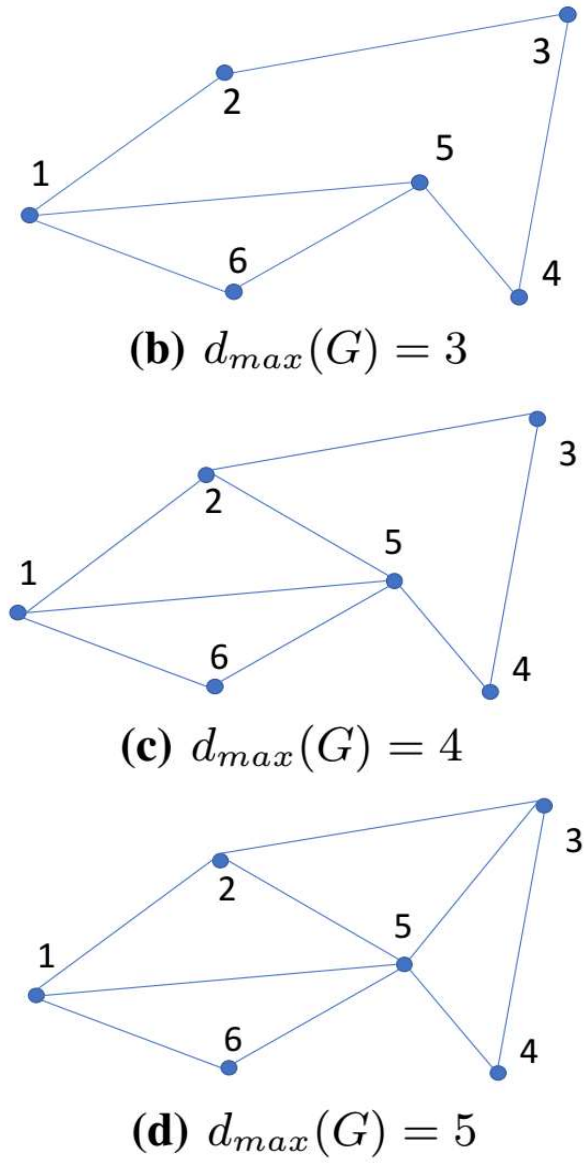}}}\hfill
\caption{(a) depicts the recognition accuracy of MSN with different topologies. (b)-(d) are the practical topologies of $d_{max}(G)=3$, $d_{max}(G)=4$, and $d_{max}(G)=5$, respectively.}
\label{fig_sim_scores}
\end{figure}

Fig.~\ref{MSN_q_Teach_compare} illustrates the performance of the MSN with various sharing frequencies $q$. 
We can observe that the proposed MSN sharing with $q\ge 3$ outperforms the original Pointnet model in recognition accuracy. 
It shows that the proposed KD-DMS can effectively grip the knowledge of other trained models through frequent sharing. 
When $q$ is greater than $3$, the final convergence accuracy of different $q$ is the same. 
A larger $q$ can provide faster convergence. 
However, referring to Eq.~(\ref{esdfqwefuioosdayoasfaa}), large $q$ also consumes considerable communication bandwidth $C$. 
Thus, the sharing frequency $q$ must be carefully configured, simultaneously considering communication bandwidth, topology features, and application performance requirements. 

Fig.~\ref{fig_sim_asdkjajldws} depicts the recognition accuracies of different models, UAVs, and sharing frequencies.
The node degrees of node 5 and node 3 are $3$ and $2$, respectively, as shown in Fig.~\ref{KD_DMS_Topo}.
Fig.~\ref{fig_sim_asdkjajldws}.(a) provides the PSN performance of node 5 and node 3. 
The performance of node 5 outperforms that of node 3 since node 5 has a larger node degree. 
Because UAVs with additional neighbors can obtain more well-trained models, significantly improving recognition accuracy. 
However, the recognition accuracy does not increase with $q$. 
The reason is that PSN needs more learning model complexity to extract features from point cloud data, which prevents performance improvement via model sharing. 
Besides, the interchange of multiple low-performance models may result in performance degradation.

The model sharing scheme with MSN is better than with PSN in point cloud recognition in Fig.~\ref{fig_sim_asdkjajldws}.
It implies that MSN can more effectively capture the attributes of the local point cloud than PSN. 
The recognition accuracy climbs as sharing frequency $q$ increases. 
However, the recognition performance does not grow gradually with $q$ but improves by leaps.
At the end of the curves, there are two stable regions with $q\ge 3$ and $q<3$. 
If $q$ is less than $3$, insufficient model sharing makes UAVs hard to get the benefits of model sharing. 
Moreover, UAVs can fully grip the knowledge of neighbors as $q\ge 3$. 
Regardless, the recognition accuracy does not raise with $q$ when $q>3$. 
Therefore, the node degree only affects the model sharing performance when $q<3$. 
In the offered graph, over three times model sharing, the accuracy of node 5 and node 3 in Fig.~\ref{KD_DMS_Topo} can converge to the same value.

We can also observe that the model sharing scheme with the CSN model has two stable accuracies with $q\ge 3$ and $q<3$. 
However, the CSN performance is not as good as that of MSN since CSN is prone to overfitting due to its deep model structure. 
Overall, the high sharing frequency and large node degree can efficiently enhance the learning ability on different datasets.

To evaluate the impact of communication topology on model sharing, Fig.~\ref{fig_sim_scores} explores the MSN model sharing with different maximum node degrees, where the sharing frequency is $5$. 
The final stable accuracies of the three topologies are roughly similar.
Further, the topology with large $d_{max}(G)$ can converge fast to stable accuracy. 
As a result, by scaling up the maximum node degree $d_{max}(G)$ of the communication topology, the model sharing obtains a faster convergence rate, which can conduct delay-sensitive applications. 
However, the large $d_{max}(G)$ will consume more communication bandwidth according to Eq.~(\ref{qweqweo0998aifgw}). 
Thus, it requires a tradeoff setup between convergence rate and bandwidth consumption. 
\begin{figure}
\centering
     \includegraphics[width=.65\textwidth]{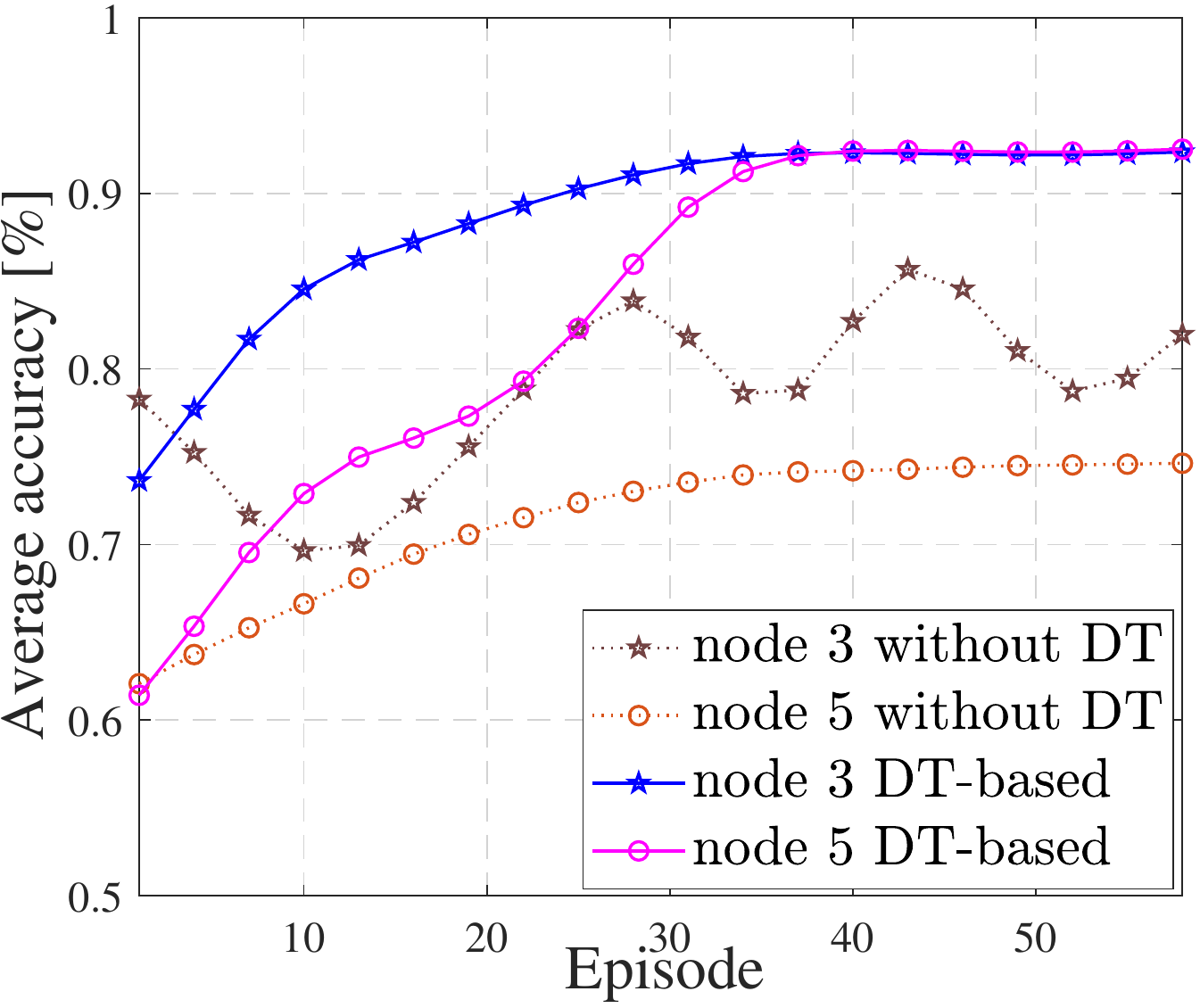} 
     \caption{Performance of DT-based model sharing versus model sharing without DT. } 
\label{loss_vs_DT}
\end{figure}

We present the recognition accuracy of model sharing with and without DT in Fig.~\ref{loss_vs_DT} to evaluate the performance gain with DT. 
The sharing topology tracks Fig.~\ref{KD_DMS_Topo}. 
While the model sharing without DT exchanges the shared model, it may cause packet loss due to the wireless transmission.
For simplification, this simulation assumes the packet loss rate of node 3 is 30\%, and node 5 is 50\% since the node with a large degree has more severe bandwidth competition. 
However, there is no transmission error in the DT asset causing the digital communication imitation.
Fig.~\ref{loss_vs_DT} reveals that DT-based model sharing has better recognition performance than model sharing without DT.
It is because the packet loss leads to inconsistent or error model sharing.
In addition, the packet loss rate of node 3 is less than that of node 5. However, node 3 only integrates two neighbor models. 
Compared with node 5, the packet loss of any neighbor in node 3 has a more unbearable impact on its performance.
It results in a significant fluctuation of the curve of node 3.

\begin{figure}
\centering
     \includegraphics[width=.65\textwidth]{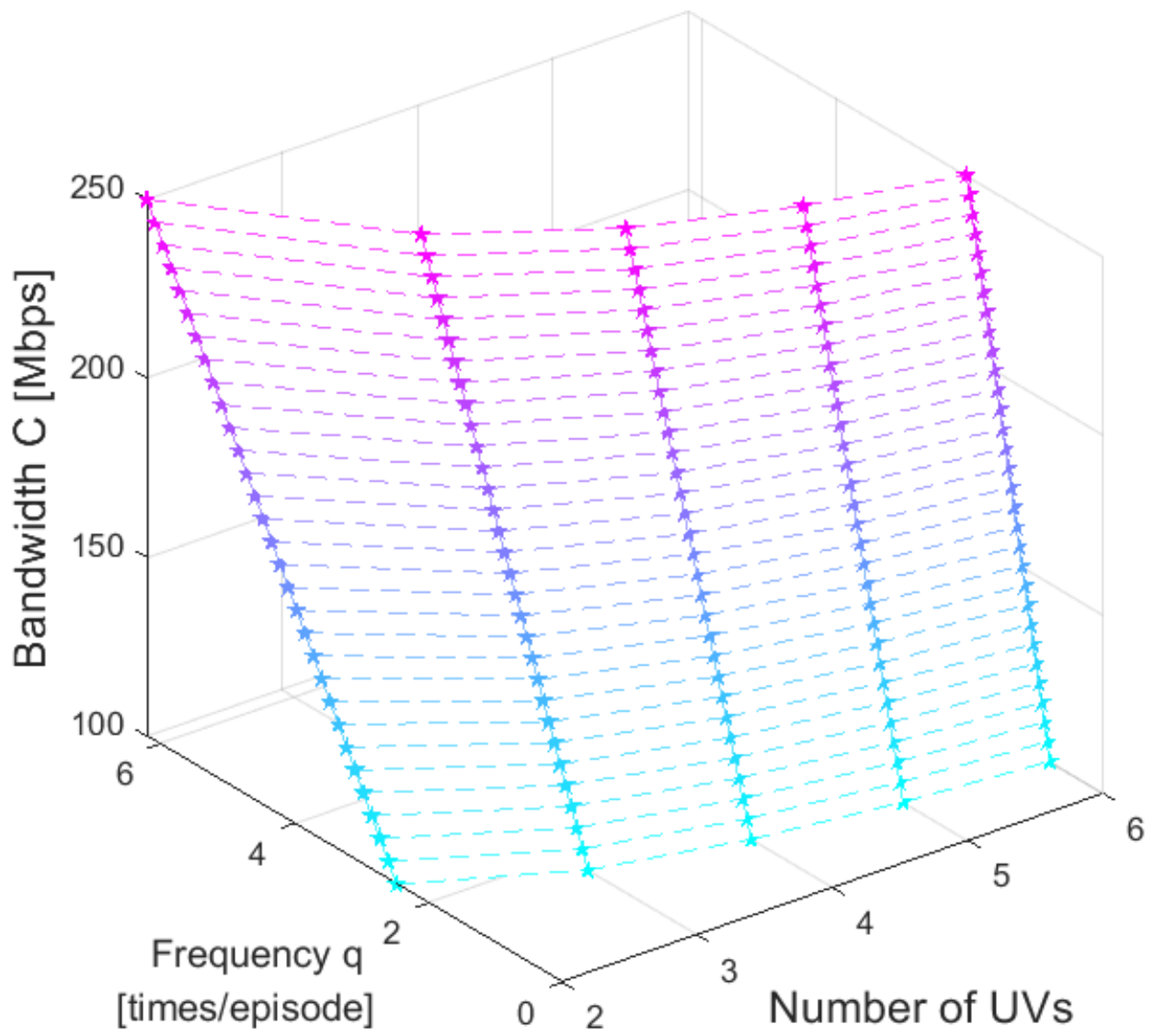} 
     \caption{Relation between bandwidth, number of UAVs, and sharing frequency. } 
\label{fre_degree_bandwidth}
\end{figure}

Fig.~\ref{fre_degree_bandwidth} depicts the relationship between the lower bound of bandwidth consumption $C$, the number of UAVs $N$, and the sharing frequency $q$, where $q$ and $N$ only take integer values. 
The frequency $q$ is proportional to bandwidth $C$ but inversely to $N$. 
As the number of UAVs $N$ scales up, the available bandwidth $C$ will decrease according to Eq.~(\ref{qweqweo0998aifgw}). 
The large $N$ also yields a small $q$ according to Eq.~(\ref{esdfqwefuioosdayoasfaa}). 
$q$ thereby is a relatively small number and unlikely to exceed $3$ when $N>5$ in most cases. 
Combining with Fig.~\ref{fig_sim_asdkjajldws}, we can predetermine the appropriate lightweight model, UAV sharing frequency, and bandwidth resource provisioning in a practical model sharing scenario.

\section{Conclusion}
This paper proposed a DT-empowered distributed model sharing scheme for 3-D point cloud recognition. 
It can be applied to resource-limited UAVs for perceiving production environments.
With the support of the device-edge DT system, the UAVs can be surveyed with timely status by the device DT; and optimized with reliable swarm imitation by the edge DT. 
Specifically, the edge DT first generates three lightweight models simultaneously through the KD compression on the edge server.
Based on the KD lightweight model, a model sharing scheme is proposed to ensure that the shared models of UAVs converge to the same model. 
The model sharing scheme leverages the network calculus to acquire the suitable shared model, network topology, and resource assignment for multiple UAV collaborations. 
Moreover, this network calculus analysis in the edge DT relies on the summarized information provided by the device DTs.
Experiment results demonstrated that the proposed model sharing scheme is feasible and outperformed the Pointnet without model sharing, in terms of recognition accuracy.
This work provisions the possibility of utilizing the collaboration of multiple resource-limited devices to improve embedded learning performance. 
It shows great potential for large-scale swarm intelligence in IIoT scenarios.


\appendices
\footnotesize
\bibliography{biblio}
\end{document}